\documentclass[useAMS,usenatbib]{mn2e}
\usepackage{rotating}
\usepackage{subfigure}
\RequirePackage{setspace}
\usepackage{graphicx}
\usepackage{fancyhdr}
\usepackage{natbib}
\usepackage{amsmath}
\usepackage{amssymb}
\usepackage{aas_macros}
\usepackage{epstopdf}

\title[Massive High-$z$ Galaxies using DES]{Detecting Massive Galaxies at High Redshift using the Dark Energy Survey}
\author[L. J. M. Davies et. al.]{L. J. M. Davies \thanks{E-mail:
Luke.Davies@bristol.ac.uk}$^{1,2}$, C. Maraston$^{1}$, D. Thomas$^{1}$, D. Capozzi$^{1}$, R. H. Wechsler$^{3}$, 
\newauthor M. T. Busha$^{4,5}$, M. Banerji$^{6,11}$, F. Ostrovski$^{7,8}$,  C. Papovich$^{9}$, B. X. Santiago$^{8,10}$,   
\newauthor  R. Nichol$^{1}$, M. A. G. Maia$^{7,8}$, L. N. da Costa$^{7,8}$\\
$^{1}$Institute of Cosmology and Gravitation, University of Portsmouth, Dennis Sciama Building, Burnaby Road, Portsmouth PO1 3FX\\
$^{2}$Department of Physics, University of Bristol, H.H. Wills Physics Laboratory, Tyndall Avenue, Bristol, BS8 1TL, UK\\
$^{3}$Kavli Institute for Particle Astrophysics and Cosmology; Physics Department, Stanford University; Department of Particle Physics  \\ and Astrophysics, SLAC
National Accelerator Laboratory; Stanford, CA 94305\\
$^{4}$Institute for Theoretical Physics, University of Z\"{u}rich, Z\"{u}rich, Switzerland\\
$^{5}$Physics Division, Lawrence Berkeley National Laboratory, Berkeley, CA 94720, USA\\
$^{6}$ Institute of Astronomy, University of Cambridge, Madingley Road, Cambridge CB3 0HA, UK\\
$^{7}$Observat\'{o}rio Nacional, Rua General Jos\'{e} Cristino, 77, 20921-400 S\~{a}o Crist\'{o}v\~{a}o, Rio de Janeiro, RJ, Brazil\\
$^{8}$Laborat\'{o}rio Nacional de e-Astronomia, Rua General Jos\'{e} Cristino, 77, 20921-400 S\~{a}o Crist\'{o}v\~{a}o, Rio de Janeiro, RJ, Brazil\\
$^{9}$ George P. and Cynthia Woods Mitchell Institute for Fundamental Physics and Astronomy, and Department of Physics and Astronomy, \\ Texas A\&M University, College Station, TX, 77843- 4242, USA\\
$^{10}$ Instituto de F\'{\i}sica, Universidade Federal do Rio Grande do Sul, Av. Bento Gon\c{c}alves 9500, CP 15051, Porto Alegre, Brazil\\
$^{11}$ Department of Physics \& Astronomy, University College London, Gower Street, London, WC1E 6BT, UK}

\begin{document}

\date{Accepted: June 2013}

\pagerange{\pageref{firstpage}--\pageref{lastpage}} \pubyear{2010}

\maketitle

\begin{abstract}

The Dark Energy Survey (DES) will be unprecedented in its ability to probe exceptionally large cosmic volumes to relatively faint optical limits. Primarily designed for the study of comparatively low redshift ($z\,<\,2$) galaxies with the aim of constraining dark energy, an intriguing byproduct of the survey will be the identification of massive ($>10^{12.0}$M$_{\odot}$) galaxies at $z\,\gtrsim\,4$. This will greatly improve our understanding of how galaxies form and evolve. By both passively evolving the low redshift mass function and extrapolating the observed high redshift mass function, we find that such galaxies should be rare but nonetheless present at early times, with predicted number densities of $\sim0.02$\, deg$^{-2}$. The unique combination of depth and coverage that DES provides will allow the identification of such galaxies should they exist - potentially identifying hundreds of such sources. We then model possible high redshift galaxies and determine their detectability using the DES filter sets and depths. We model sources with a broad range stellar properties and find that for these galaxies to be detected they must be either sufficiently young, high mass and/or relatively dust free (E(B-V)$<$0.45) - with these parameters jointly affecting each galaxy's detectability. We also propose colour-colour selection criteria for the identification of both pristine and dusty sources and find that, although contamination fractions will be high, the most reliable candidate massive high redshift galaxies are likely to be identifiable in the DES data through prioritisation of colour-selected sources.

\end{abstract}


\section{Introduction}
\label{sec:intro} 

With the impressive advancement in astronomical technologies over the last decade, the boundaries of high redshift ($z$)  astrophysics have been pushed further and further into the early Universe. Now, using modern telescopes and a variety of techniques, large samples of galaxies have been observed back to within $\sim1\,$Gyr of the Big Bang \citep[$e.g.$][]{Stanway04,Vanzella09, Douglas09, Douglas10, Bouwens10, Bunker10, Finkelstein10}. Through optical imaging, analysis of galaxy spectra and spectral energy distribution fitting, a wealth of information has been uncovered about these sources, allowing observations of galaxy formation and evolution at its very earliest stages. The stellar mass function (SMF) of galaxies has now been measured out to $z\sim7$ \citep[e.g.][]{Dickinson03,Conselice05a, Perezgonzalez08, Bundy09, Marchesini09, Stark09, McLure09, Banerji10}, providing estimates of the stellar mass of the Universe when it was in its infancy.  

However, the large investments of telescope time which are required to identify high redshift sources have limited observations to the host galaxies of bright energetic events \citep[such as quasars and gamma ray bursts, \textit{e.g.}][]{Akiyama05, Conselice05} or faint galaxies in small area fields \citep[\textit{e.g.}][]{Douglas09, Douglas10}. This has restricted most studies to sources which are either atypically bright (hence can be detected in shallow, large scale surveys) or have high number densities (therefore can be detected in small fields). 

Hence, most previous studies could be missing an important subset of high redshift galaxies, namely rarer, massive galaxies ($>10^{12}\,$M$_{\odot}$) which may be already passively evolving at $z\sim3$.

This class of galaxies is crucial to our understanding of galaxy formation and evolution within a cosmological framework. In fact, most models of galaxy formation in the literature predict that the most massive objects complete their mass assembly at low redshifts \citep[\textit{e.g}][]{Delucia06, Ricciardelli10} and hence a paucity of such massive objects at high redshift. On the other hand, studies of the fossil stellar population record in the local Universe show that the most massive galaxies are those hosting the oldest stellar populations and should have formed at $z\sim5$, the well-known {\it downsizing} paradigm of galaxy evolution \citep[][$etc$]{Cowie99, Kuntschner01, Nelan05,Thomas05, Panter07, Cowie08, Conroy09, Thomas10}. It yet remains to be seen whether the stars present in massive elliptical galaxies formed in small systems at high redshift (over a brief period of common starburst activity) which later merged \citep[\textit{e.g}][]{Delucia06}, or formed (and rapidly built-up) in massive systems which have passively evolved since. 

Studies of high-redshift star forming galaxies do identify large samples of small galaxies which are undergoing rapid, relatively short-lived star formation episodes \citep[$e.g.$][]{Verma07}, and hence may form the building blocks of current day massive galaxies \citep[\textit{e.g.}][]{Giavalisco96}. However, the stellar mass contained within over-densities of these systems is not large enough to produce a giant elliptical at the current epoch  \citep[\textit{e.g.}][]{Douglas10}. Although we may only observe a small fraction of systems which may eventually form stars in any particular survey volume, the small range for star formation timescales observed in local massive galaxies impose tight constraints on their formation epochs \citep{Thomas05}. If giant elliptical galaxies formed their stars over a brief $\sim$1\,Gyr period at $z\sim5$, then we should see enough star formation activity in small systems at $z\sim5$ (and in small enough regions) to produce the stellar masses observed in massive low redshift galaxies. As yet this is not the case \cite[$e.g.$][]{Douglas10} however, the potential sites of massive galaxy formation will be rare and hence may have avoided small area high redshift surveys. 

In a converse approach the detection of rare, rest-frame UV-bright, massive star forming systems at high redshift \citep[which may passively evolve to giant elliptical galaxies seen in the local universe, $e.g.$][]{Wake06} would constrain current galaxy formation models. Recent studies have identified a small number of massive (M$_{\mathrm{star}} > 2.5 \times 10^{11}$M$_{\odot}$) galaxies at $z\sim3-4$ \citep[$e.g.$][]{Papovich06,Marchesini10}, with potential discoveries of such systems at $z\gtrsim5$ \citep[$e.g.$][]{Yan06, Mancini09}. However, the evidence for massive, evolved galaxies at $z>4$ is still unconvincing  \citep[e.g.][]{Dunlop07}.

The next generation of large galaxy surveys (\textit{e.g.} the Dark Energy survey, DES), in conjunction with spectroscopic follow up, will provide a unique opportunity to detect these rare, bright galaxies at the highest redshifts ($z\,>\,4$) - should they exist. DES will generate data that will be two to three magnitudes deeper than existing large surveys (\textit{i.e.} the Sloan Digital Sky Survey) and will cover $\sim32$ thousand times more area than current small area high redshift surveys. DES will probe an unprecedentedly large volume enabling the detection of the much rarer, bright (and hence massive) systems which may have evaded smaller area surveys.

In this work, we outline possible sources that could potentially be detected by DES using model galaxy spectra and develop colour selection criteria for their identification. We also make predictions of the number counts of massive galaxies likely to be detected by DES from current galaxy formation models and by evolving the local mass function. Throughout this paper we uses cosmological parameters: $\Omega_{m_{0}}=0.25$, $\Omega_{\Lambda_{0}}=0.75$ and $H_{0}=75$ km\,s$^{-1}$\,Mpc$^{-1}$. All magnitudes are given in the AB$_{v}$ system \citep{Oke83}. Colour versions of all figures in this paper are available in the online version.

\section{The Dark Energy Survey}

The Dark Energy Survey is a large optical imaging program designed to cover an exceptionally large are over the southern Galactic pole. The survey is split into a wide area survey (DES-W) covering 5000\,deg$^{2}$ and reaching a depth of $\sim25.5, 25.0, 24.4, 23.9$ and 22.0 \,mag in g, r, i, z and Y bands respectively \citep[for a 5$\sigma$ detection, ][]{Rossetto11}, and a deep survey (DES-D) covering 30\,deg$^{2}$ in two `deep' and eight `shallow' fields and reaching depths of $\sim27.1, 27.3, 27.0, 26.8$ and 21.7 \,mag in g, r, i, and z bands respectively in the `deep' fields \citep[for a 5$\sigma$ detection, see ][ for further details]{Bernstein12}. The main survey region has been carefully designed to overlap with the South Pole Telescope (SPT) and the Vista Hemisphere Survey (VHS) to provide partial coverage at additional wavelengths. Observations will be carried out by the specifically commissioned Dark Energy Camera (DECam), mounted on the Blanco 4m telescope at the Cerro Tololo Inter-American Observatory (CTIO) in La Serena, Chile. DECam is an array of 64, 2048\,$\times$\,4096 CCDs designed to to be optimally efficient at 9000\AA, covering a wavelength of 4000-10000\AA.  It provides a 2.2 deg$^2$ field of view with a 0.27$^{\prime\prime}$\,pixel$^{-1}$ resolution; the instrument providing $<\,0.3^{\prime\prime}$ FWHM to the resultant image quality \citep{Banerji08}. 

The main scientific goal of the project is to investigate dark energy by measuring cosmic expansion via baryon acoustic oscillations, galaxy clustering, weak lensing and supernovae. However, the combination of survey area and depth also make DES an invaluable data set for galaxy evolution studies. In particular DES will allow us to search for rare objects and events such as massive galaxies and mergers or nearby objects which require a combination of depth and large area coverage \citep[such as Milky Way satellites, see][]{Rossetto11} . 
 
  \begin{figure*}
\begin{center}
\includegraphics[scale=0.50]{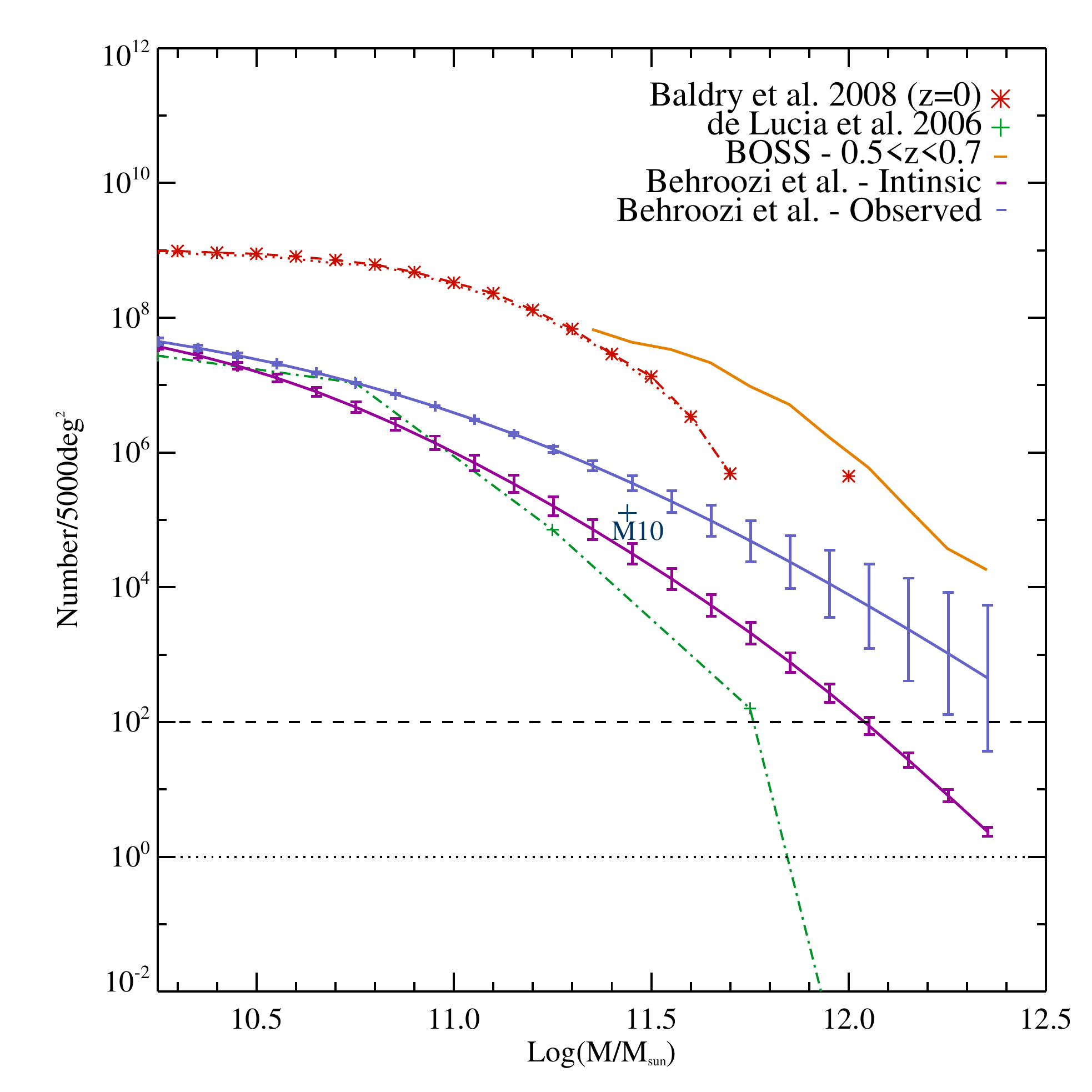}
\caption{Predicted number counts of galaxies in the redshift range $3\,<\,z\,<\,6$ for 5000\,deg$^{2}$ (area covered by DES). Green dot-dashed line displays the predicted number counts from \citet{Delucia06} galaxy formation models in a $\Lambda$CDM universe (represented by the Millennium Simulation). The number counts in the log$_{10}$(M$_{*}$/M$_{\odot}$)$\,<\,11.5$ region are taken directly from the simulation mass function (where cosmic variance on scales larger then the simulation volume will not affect the number counts), while the log$_{10}$(M$_{*}$/M$_{\odot}$)$\,=\,12.25$ point (well below the Y-range of the plot) is an extrapolation from the lower mass numbers. The dashed red line displays the predicted number counts produced by assuming passive evolution from the $z\,=\,0$ mass function from \citet{Baldry04}. The solid orange line displays the BOSS mass function for $0.5<z<0.7$ galaxies (Maraston et al. 2012). The solid purple lines display the summary of the predicted $3<z<6$ mass function taken from \citet{Behroozi12} which is fit to a compilation of existing data in the literature. We display both the observed (including Eddington bias) and intrinsic mass function. Hence, the observed mass function displays the number of galaxies which may be identified in the DES survey volume, while only the intrinsic number of galaxies will actually be high mass systems.  The dotted black line represents one source, while the dashed black line represents 100 sources. In the de Lucia model,  less than one log$_{10}$(M$_{*}$/M$_{\odot}$)$\,>\,12.0$ source is predicted (by extrapolating from lower masses), while passive evolution models predict thousands of galaxies - providing an upper and lower limit. The number counts of M$_{\mathrm{star}} > 2.5 \times 10^{11}$M$_{\odot}$ galaxies identified at $z\sim3-4$ in the NEWFIRM Medium-Band Survey \citep[NMBS][]{Marchesini10}, scaled to the DES survey area is displayed as the blue cross marked M10.}
\label{fig:num_predict}
\end{center}
\end{figure*}

\section{Predicted Number Counts of Massive Galaxies}
\label{sec:nums}

In order to make predictions of the possible numbers of massive high redshift galaxies which are likely to be detected by DES, we compare the predicted numbers of sources in the DES-W survey volume (the largest volume probed by the two DES surveys) for a hierarchical growth model (as traced by the Millennium Simulation), by passively evolving the $z=0$ and $0.5<z<0.7$ mass functions, and using the $3<z<6$ mass function from \citep{Behroozi12}, based on a fit to existing data in the literature. The extreme cases of the hierarchical growth model and passive stellar evolution provide a lower and upper limit respectively on the number of galaxies that DES may identify, while the \cite{Behroozi12} values display a more realistic prediction of the true number of sources. Hierarchical growth models predict a paucity of the most massive objects at high redshift, while passively evolving the lower redshift mass functions assumes all low redshift giant ellipticals formed at high redshift (thus providing an upper limit).            

The Millennium Run simulation \citep{Springel05} uses $\Lambda$CDM cosmology to trace the evolution of cold dark matter (CDM) in the universe via N-body simulations and delineates the gravitational buildup of large scale structure via hierarchical growth. It follows a CDM cosmology (parameters outlined in section \ref{sec:intro}) and is congruous with the WMAP year 1 data \citep{Spergel03}. Semi analytic galaxy formation models \citep[\textit{e.g.}][]{Hatton03, Delucia06, DeLucia07, Monaco07, Menci12} are applied to the CDM framework to investigate the evolution of baryonic matter in a $\Lambda$CDM universe. Hence, using these models we can make predictions about the likely number densities of galaxies at any given epoch within the $\Lambda$CDM framework. We use the de Lucia et al. (2006) semi-analytic galaxy formation model run on the Millennium simulation to produce a mock pencil beam pointing between $z=3-6$ and determine the mass function for luminous sources of $>10^{10}$M$_{\odot}$. The volume of this pointing is scaled to the size of the full DES simulation volume (5000\,deg$^{2}$). 

No log$_{10}$(M$_{*}$/M$_{\odot}$)$\,>\,12.0$ galaxies are found within the simulation. However, the lack of these most massive systems is likely to be an artefact of simulation volume being inadequate in containing the most massive systems (the simulation does not trace the full spectrum of cosmic variance in a hierarchical model). We therefore extrapolate the mass function from lower mass galaxies using a Schechter function fit. We note that while the Millennium Run simulation value of $\sigma_{8}$ may not be consistent with the WMAP7 data, applying the more recent value of $\sigma_{8}$ would not significantly change the predictions made in this paper - it suggests lower numbers counts of massive high redshift sources in a $\Lambda$CDM Universe.

In contrast, using a passive stellar evolution model the co-moving number density of massive galaxies at high redshift is forced to be equal with those at low redshift. \cite{Baldry04} used SDSS to calculate the low redshift stellar mass function for early type galaxies assuming a \cite{Kroupa01} type IMF. We calculate the co-moving volume of DES at $3\,<\,z\,<\,6$ and determine the predicted numbers of galaxies assuming this passive evolution. We also calculate the predicted numbers of massive high redshift galaxies by passively evolving the Baryon Oscillation Spectroscopic Survey (BOSS) mass function for $0.5<z<0.7$ sources, for  a \cite{Kroupa01} type IMF \citep{Maraston12}.  As discussed previously, these passive evolution estimates provide an upper limit to the possible number of galaxies DES may identify.

Lastly, in order to produce a more realistic estimate for the true number of massive high redshift sources we use the recent summary of the high redshift mass function taken from \cite{Behroozi12} - using a \cite{Chabrier03} IMF.

\begin{figure*}
\begin{center}
\includegraphics[scale=0.50]{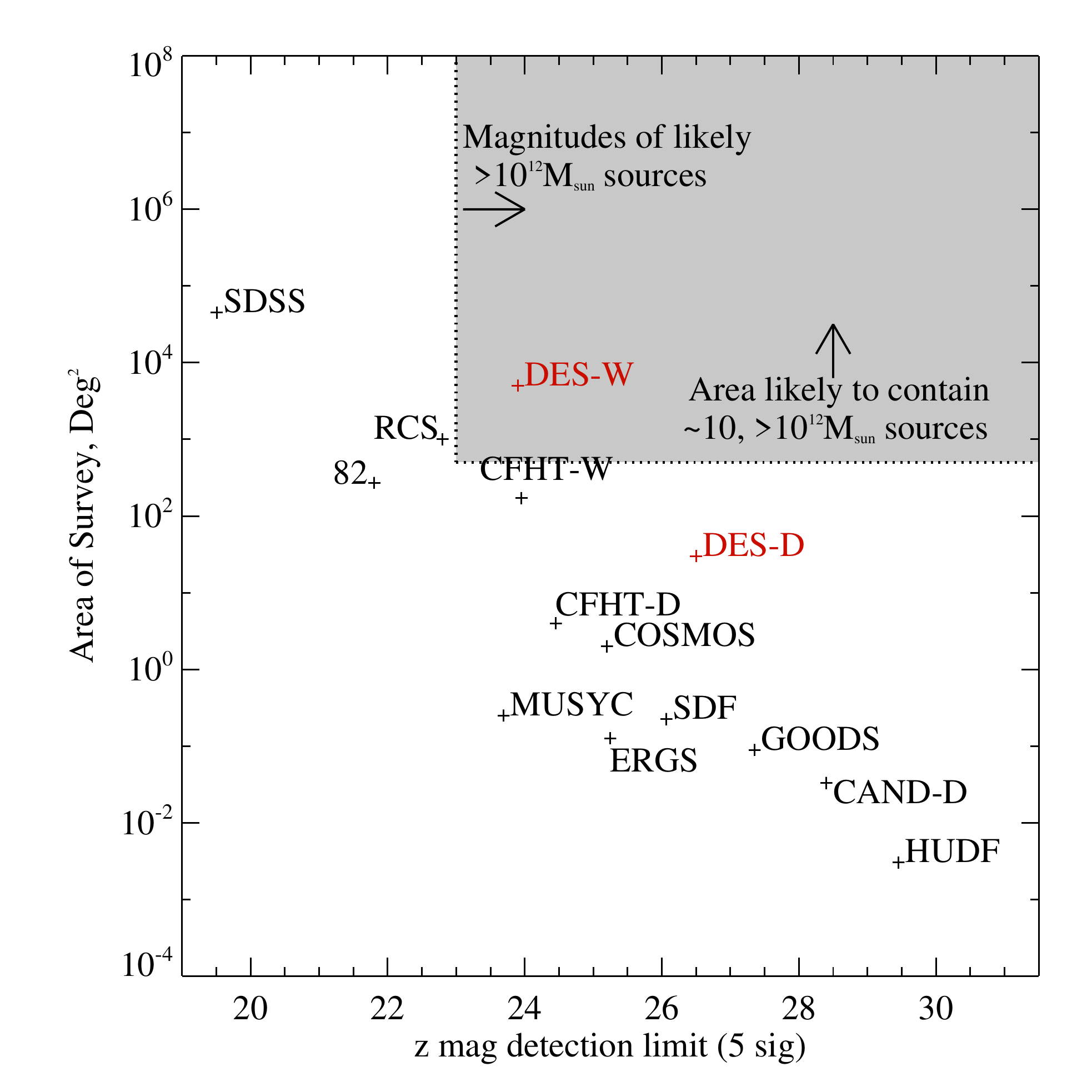}
\caption{A comparison of z-band magnitude limit against area for a sample of surveys. Grey region represent the area of parameter space which is likely to detect M$\,>\,10^{12}$M$_{\odot}$ galaxies at $z\,>\,3$. DES-W falls within this region as it covers a large area and is relatively deep in Z. However, we find that the DES-D survey does not cover sufficient area in order to detect these rare sources. Other surveys plotted: Sloan Digital Sky Survey DR7 \citep[SDSS,][]{Abazajian09}, The Red-sequence Cluster Survey-2 \citep[RCS,][]{Gilbank10}, SDSS stripe 82 \citep[82, $e.g.$][]{Frieman08}, The CFHT Legacy Survey - Wide (CFHT-W),  The CFHT Legacy Survey - Deep (CFHT-D), The ESO Remote Galaxy Survey \citep[ERGS,][]{Douglas09}, The Subaru Deep Field \citep[SDF,][]{Kashikawa04}, The Cosmic Assembly Near-infrared Deep Extragalactic Legacy Survey (CANDELS) - deep \citep[CAND-D,][]{Grogin11,Koekemoer11} - note that CANDELS-wide is not included as it does not have z-band coverage, The Great Observatories Origins Deep Survey \citep[GOODS,][]{Giavalisco04}, Hubble Ultra Deep Field \citep[HUDF,][]{Beckwith06}, MUltiwavlength Survey by Yale-Chile ECDF-S field \citep[MUSYC,][]{Gawiser06}, The Cosmic Evolution Survey \citep[COSMOS,][]{Capak07} - note, the COSMOS area is for the ground-based z-band data.}
\label{fig:surveys}
\end{center}
\end{figure*}    

Figure \ref{fig:num_predict} displays the predicted number counts of galaxies in a DES-W survey area at $3<z<6$ as a function of mass, for a hierarchical growth model (green dot-dashed line), from passively evolving the $z=0$ (red dashed line) and $0.5<z<0.7$ (solid orange line) mass functions and for the $3<z<6$ mass function of  \cite{Behroozi12} (purple lines). The de Lucia model predicts less than one log$_{10}$(M$_{*}$/M$_{\odot}$)$\,>\,12.0$ galaxy at $z>3$ in comparison to an unrealistic maximum of $\sim100,000$ sources predicted from passively evolving the lower redshift mass functions. Clearly, the identification of any log$_{10}$(M$_{*}$/M$_{\odot}$)$\,>\,12.0$ source at $z>3$ in DES will help constrain galaxy formation models. The \cite{Behroozi12} mass function predicts that $\sim7000$ log$_{10}$(M$_{*}$/M$_{\odot}$)$\,>\,12.0$ galaxies should be observed in the DES survey volume. However, only $\sim100$ of these would represent true log$_{10}$(M$_{*}$/M$_{\odot}$)$\,>\,12.0$ systems, with the rest being identified as massive galaxies through Eddington bias \citep[$e.g.$][]{Li09, Maraston12}. This number of sources predicted from \cite{Behroozi12} represents the best estimate for the true number of massive sources at $z>3$. Hence, in our further analysis we will assume this number density.          

For comparison we also display the number of  M$_{\mathrm{star}} > 2.5 \times 10^{11}$M$_{\odot}$ galaxies identified at $z\sim3-4$ in the NEWFIRM Medium-Band Survey \citep[NMBS,][]{VanDokkum09, Marchesini10}, scaled to the DES survey area. This value is consistent with the true number of massive high redshift galaxies falling between the semi-analytic and passive evolution predictions and close to the  \cite{Behroozi12} predictions. While the \cite{Marchesini10} findings help constrain the number density of massive high redshift galaxies, the small area probed ($\sim0.5$\,deg$^2$) does not allow the identification of more massive (and hence rarer) systems which may be identified in DES. However, if we use the best fit Schecter function parameters obtained in \cite{Marchesini10} for both \cite{Bruzual03} and \cite{Maraston05} stellar population models, and extrapolate to higher masses, we find $\sim$1.5 and 0.02 $>10^{12}$M$_{\odot}$ systems in the DES survey volume respectively - much lower than those predicted by \cite{Behroozi12}.                       
 
However the question remains, if such systems exist should they have already been identified in current surveys? The predicted number counts of true sources from \cite{Behroozi12} predict $\sim1$\, Log$_{10}$(M$_{*}$/M$_{\odot}$)$\,>\,12.0$ source per 50\,deg$^{-2}$. While some large scale surveys cover sufficient area to identify such sources ($e.g.$ SDSS), the do not have the required depth. In order to produce the stellar masses of giant ellipticals over a brief period of time, we might expect such systems to be `scaled up' versions of a more normal high redshift star-forming galaxy. Typical lyman-break galaxies at $z\sim5$ have masses $\sim10^{9.5}$\,M$_{\odot}$ and have z-band magnitudes of $\sim25.5$ \citep{Verma07}. Assuming a constant stellar mass to light ratio ($i.e.$ the luminosity scales directly with mass), a $\sim10^{12}$\,M$_{\odot}$ source at $z\sim5$ would potentially have a z-band magnitude of $>\,23$. This may not be the case as the observable UV-bright stellar populations in LBGs may not represent their true stellar mass. Underlying, older stellar populations (from previous star formation episodes) may form a large fraction of their stellar mass \citep[$e.g.$][]{Pentericcci09, Maraston10}. Observations of these underlying populations are problematic at all wavelengths due to \textit{overshining} by the younger component. However, due to the lack of detailed understanding of older stellar populations in LBGs at $z\sim5$, in this simple analysis we assume the mass derived from the UV-bright component is representative of the true mass of the galaxy.      

Figure \ref{fig:surveys} displays the $5\sigma$ z-band depth against area for a sample of current surveys. Shallow, large area surveys (such as SDSS or RCS) fall at the top left of the plot and deep small field of view surveys (such as SDF and HUDF) to the bottom right. The grey shaded region represents the parameter space in which a survey is likely to detect  $\gtrsim10$ massive, high redshift galaxies - taken from the predicted number counts assuming the \cite{Behroozi12} mass function and `scaling up' the flux from a typical high redshift star-forming galaxy (as described above). All existing wide area surveys are too shallow and deep surveys too small in area to identify these massive high redshift galaxies. However, the unique combination of depth and area coverage of DES-W lies above the general trend of decreasing z-band limit with area and thus will provide the first opportunity to detect these galaxies, should they exist. We find that, although above the general trend of deceasing area with survey sensitivity, the DES-D survey does not cover sufficient area to be likely to detect these highly rare sources. Hence, in any further discussion we shall only considier the DES-W survey.

  %%%%%%%%%%%%%%%%%%          

\section{Modelling Massive Galaxies}
\label{sec:models}

Due to its large volume and depth, DES will contain over 300 million galaxies. Thus, differentiating massive high redshift galaxies from their smaller counterparts at lower redshift will be the major challenge. We therefore model possible massive high redshift galaxies and determine their detectability and possible colours using the DES filters.

\begin{figure*}
\begin{center}
\includegraphics[scale=0.65]{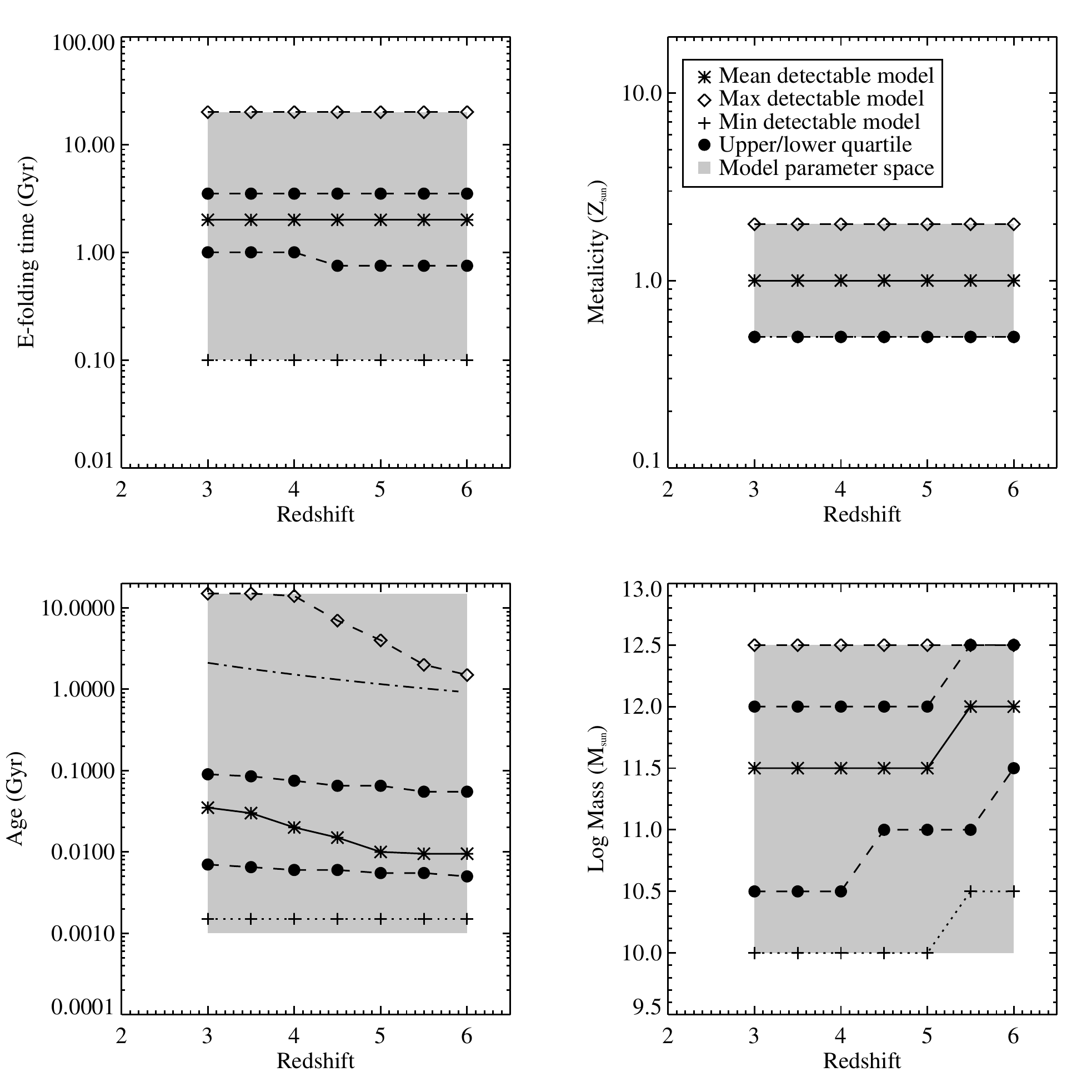}
\caption{The trends of mean e-folding time, metallicity, age and mass with redshift of model galaxies likely to be detected at the DES $5\sigma$ limiting magnitudes in one of the DES filters. Solid line with stars represents the mean values, dashed line with diamonds represents the maximum values, dotted line with crosses represents the minimum values and dashed line with circles represents the upper and lower quartiles of the distribution. Grey regions display the parameter space probed using all stellar population models. The e-folding time and metallicity of detectable sources cover the full input parameter space and do not vary with redshift. Hence they have little affect on the detectability of our sources. As we move to higher redshift, galaxies that may be detected become younger and more massive. DES is likely to detect galaxies with total stellar mass log$_{10}$(M$_{*}$/M$_{\odot})\,>\,10.0$ at $z\,>\,3$ and galaxies of log$_{10}$(M$_{*}$/M$_{\odot})\,>\,10.5$ at $z>5$. However, the typical detected galaxy is likely to be young ($\sim10-50$\,Myr) and massive ($>10^{11.5}$M$_{\odot}$). Bottom left: Dot-dashed line displays the age of the Universe at each redshift. In this figure we plot all possible models, but we later discard those that are older then the age of the Universe at a given redshift.}
\label{fig:trends}
\end{center}
\end{figure*}

\subsection{Stellar Population Templates}

\label{sec:SPM}

Using a library of stellar population models \citep{Maraston05} we model massive galaxies at high redshift.  We consider a broad range of potential star-formation histories and galaxy parameters utilising the full parameter space of the \cite{Maraston05} models.  We use simple stellar populations (SSPs) with \cite{Kroupa01} initial mass function (IMF), 1/200, 1/50, 0.5, 1.0, 2.0, 3.5 solar metallicities and ages ranging from $10^{3}$\,yrs - 15\,Gyr. We also use composite stellar population (CSPs) models with exponentially decaying star formation rates (SFR) and e-folding time-scales ranging from 0.1 to 20\,Gyr. We use CSPs with \cite{Kroupa01} IMF for solar metallicity and \cite{Salpeter55} IMF for 2, 1, 0.5 and 0.2 solar metallicity, over ages ranging from 1\,Myr to 15\,Gyr.  For both sets of models we use total stellar masses ranging from 10.0$\,<\,$log$_{10}$(M$_{*}$/M$_{\odot}$)$\,<$\,12.5 at 0.5 intervals. We also investigate the inverted tau models (with SFR $\sim e^{+t/\tau}$) of \cite{Maraston10} for solar metallicity and 0.1, 0.5 and 3 Gyr e-folding time-scales with ages between 1\,Myr to 3\,Gyr. Recent studies have shown that these types of star-formation histories may be applicable at high redshift \citep[$e.g.$][]{Papovich11}. We note that we do not consider more stochastic star-formation histories. High redshift sources are likely to be detected through their instantaneous (UV-bright) star-formation. Hence, their detectability is only likely to depend of their current burst of star-formation (and not previous star-formation episodes) - which is well represented by a single burst model.

Each stellar population model is shifted and stretched in wavelength to represent a galaxy at $z\,=$ 3, 3.5, 4, 4.5, 5, 5.5 and 6 (correcting for luminosity distance dimming and bandwidth stretching). Initially all SSP and CSP models are considered at all redshifts with no consideration of formation epoch. This may lead to the unrealistic formation redshifts for some models, however these will be removed at a later stage. All models are also corrected for neutral hydrogen absorption via the prescription outlined in \cite{Madau95}. Model galaxy fluxes are convolved with the DES filter sets  and we calculate observed magnitudes. Any model galaxy with a magnitude of brighter than the DES $5\sigma$ limiting magnitudes (25.5, 25.0, 24.4, 23.9, 22.0 in g, r, i, z and Y respectively) in any band is identified as a possible source that has the potential to be detected by DES. While this formal 5$\sigma$ limit may not represent a true 5$\sigma$ detection in the resultant data (once systematic errors are included), we note that in typical high redshift galaxy studies $\sim3\sigma$ detection limits (with additional visual classification) are applied \citep[$e.g.$][]{Douglas09}. Hence, we model sources which will be detectable at a formal 5$\sigma$ limit and note that these sources may fall below this limit in the resultant DES data.

\subsection{Detectability as a Function of Model Parameters}

Figure \ref{fig:trends} displays the trends with redshift of e-folding time, metallicity, age and mass of models that would be identified as brighter than the DES limits in any band for the CSP models. At this stage we only consider model galaxies with zero extinction, but we will discuss the effects of dust obscuration in section \ref{sec:dust}. We display the median (solid line), minimum (dotted line) and maximum (dashed line) for the parameters of detectable objects. In addition, the dot-dashed line in the bottom left panel displays the age of the Universe at each redshift, in the adopted cosmology. Since all of the mock galaxies are considered at all redshifts, independently of the age of the Universe within the adopted cosmological model, some model galaxies which are identified have ages which are older than the Universe (as discussed previously), these are disregarded in any subsequent analysis. We also note that a number of these galaxies display unrealistic star-formation histories, requiring exceptionally large SFRs in order to build their stellar mass in their young lifetimes. However, as we know nothing of the nature of these potential, massive high redshifts systems, we do not rule out an potential model at this stage of our analysis.  The full parameter space investigated using the CSP models is highlighted with the grey regions. E-folding time scale and metallicity do not vary over the large redshift range probed and cover the full range of parameter space.  Hence, these parameters will not affect the type of galaxies which are likely to be detected. 

As expected, the limits of detectable galaxies decrease in age (as depicted by the maximum age) and increase in mass (as depicted by the minimum mass) at higher redshifts as younger and more massive systems are brighter and hence better detectable at larger luminosity distances. DES is therefore likely to detect galaxies which are log$_{10}$(M$_{*}$/M$_{\odot})\,>\,10.0$ and $<$\,10\,Gyr old at $z\,>\,3$ and galaxies which are log$_{10}$(M$_{*}$/M$_{\odot})\,>\,10.5$ and $\lesssim$\,2\,Gyr old at $z\,>\,5$. 

\begin{figure*}
\begin{center}
\includegraphics[scale=0.7]{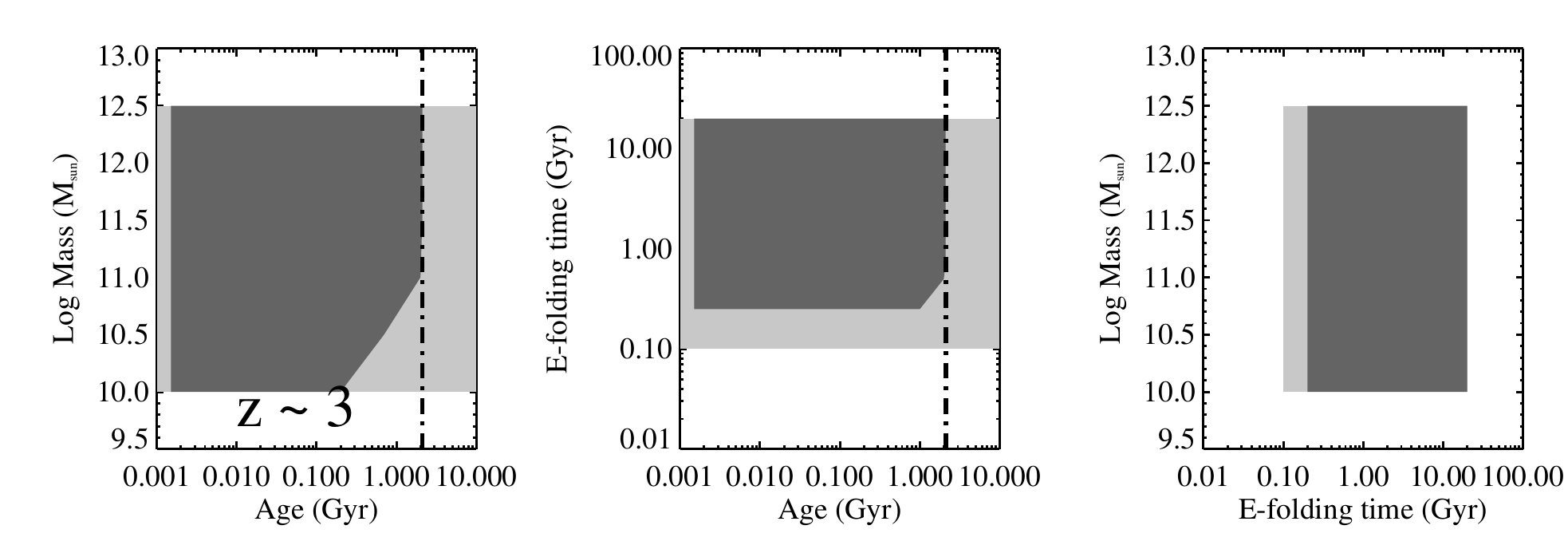}
\includegraphics[scale=0.7]{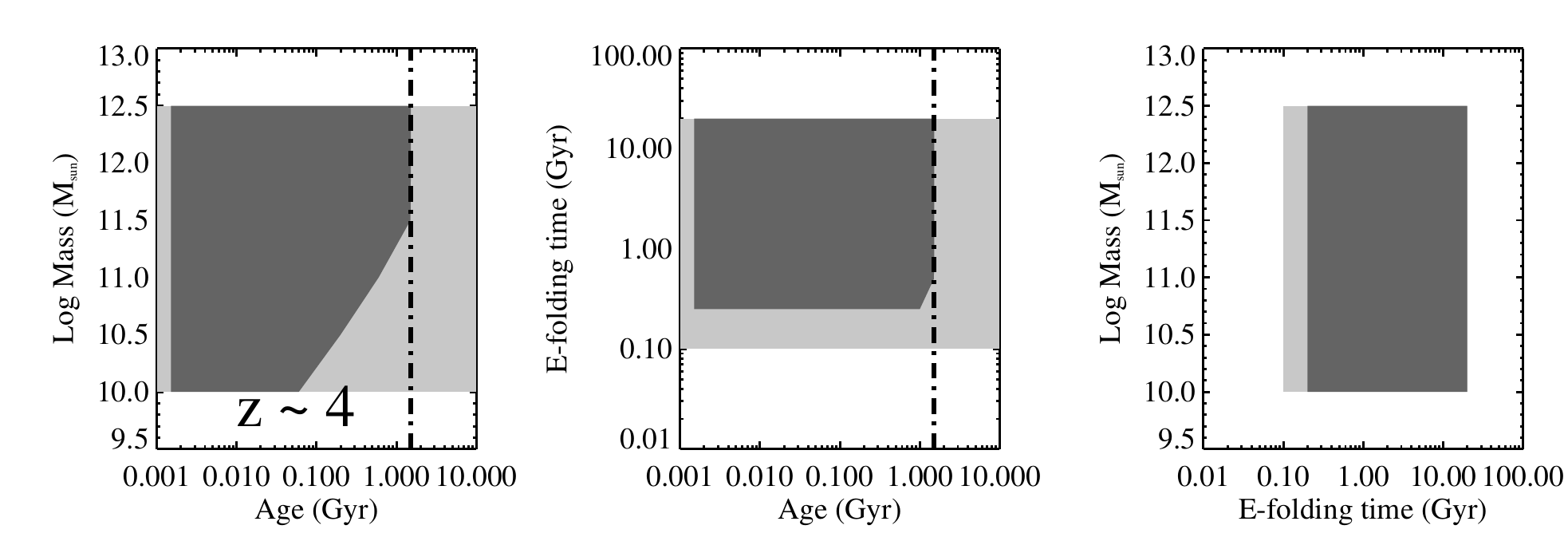}
\includegraphics[scale=0.7]{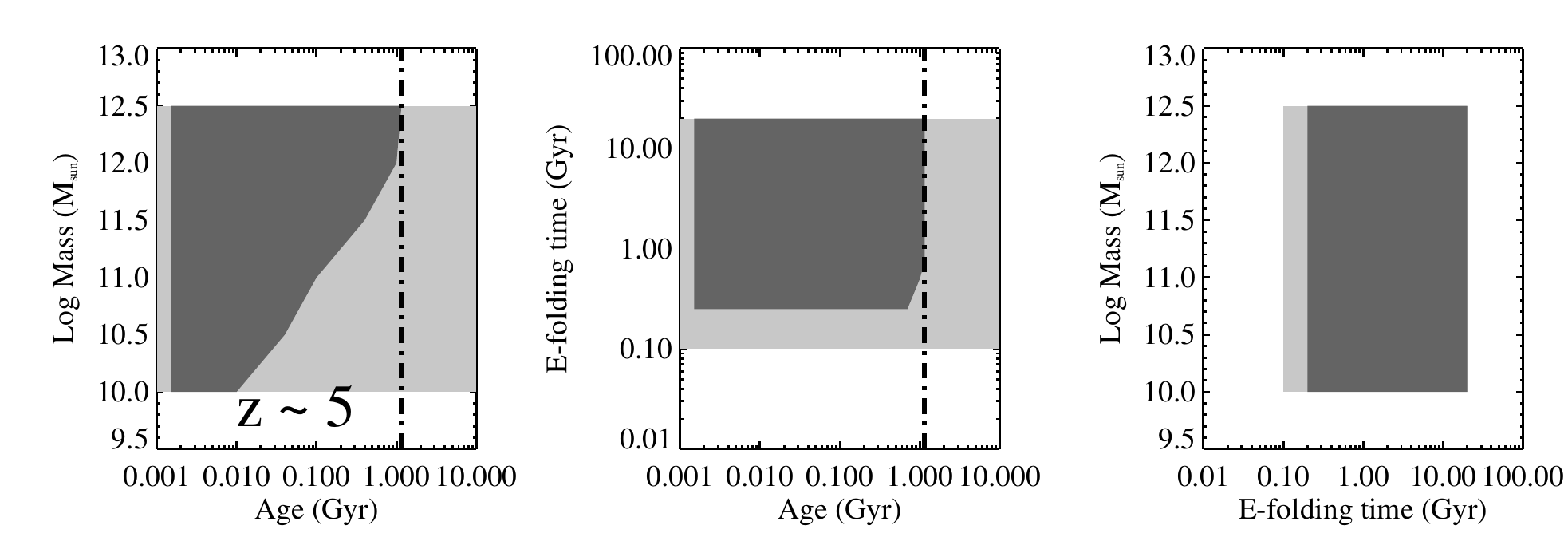}
\includegraphics[scale=0.7]{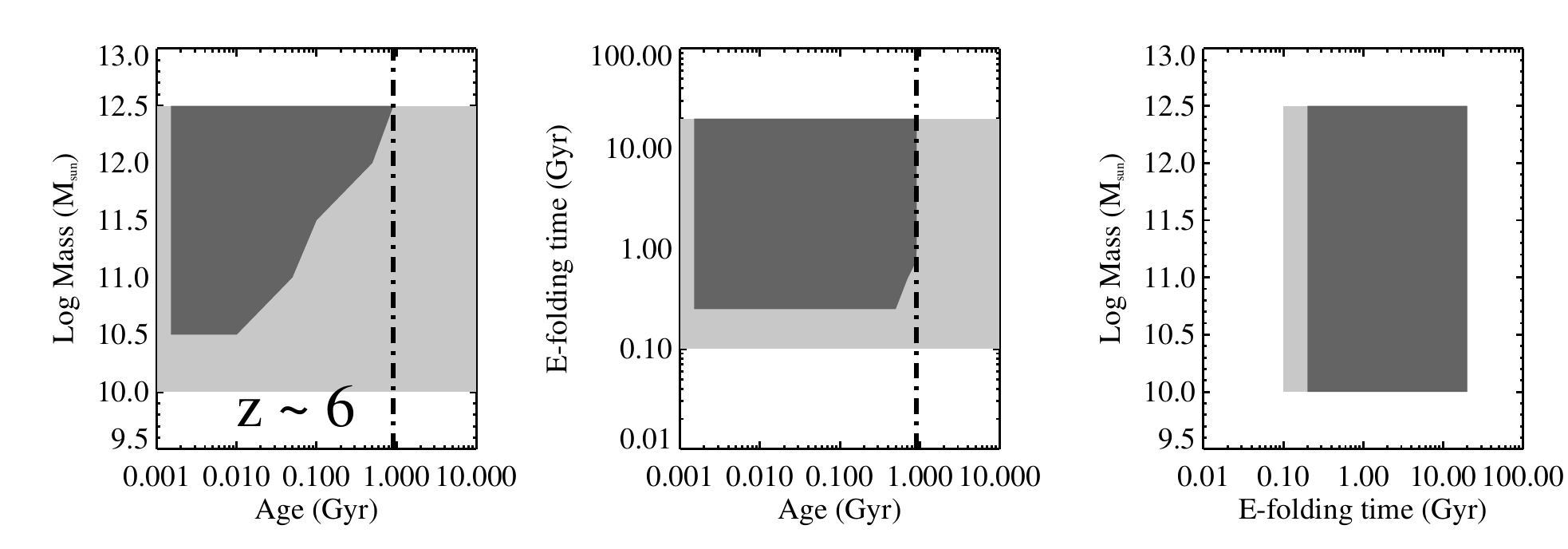}

\caption{Joint constraints of mass vs age, e-folding time vs age and mass vs e-folding time  at $z$=3, 4, 5 and 6 of objects likely to be detectable by DES. The dark shaded regions represent the parameter space covered by model galaxies which would be detected as brighter than the limiting magnitude in at least one of the DES filters.The light grey region represents the parameter space probed by the models. The dot-dashed vertical line represents the age of the Universe at each redshift. There is no joint constraint between mass and e-folding time as a full range of e-folding times are detected for all masses (right panels). At large e-folding times there is no constraint on galaxy age. However, at lower e-folding time scales a small number of galaxies must be younger to be detected (middle panels). There is also a strong constraint between mass and age as the least massive objects must also be young to be detected (left panels). For example at $10^{11}$M$_{\odot}$ galaxy at $z\sim5$ must be $< 100$Myr old to be detected. Therefore galaxies must either be young or very massive in order to be detected. }       
\label{fig:joint_trends}
\end{center}
\end{figure*}

Obviously, these galaxy parameters are not independent. Figure \ref{fig:joint_trends} displays the joint constraints of mass vs age, e-folding time vs age and mass vs e-folding time  at $z$=3, 4, 5 and 6 of objects likely to be detectable by DES. There is little joint constraint between mass and e-folding timescale as the full range of e-folding timescales are detectable at all masses. At large e-folding timescales objects have a range of ages with the upper limit governed by the redshift of the source, however at lower e-folding times a small number of galaxies must be younger in order to be detected. Obviously, there is a strong joint constraint between mass and age. At large masses our model sources once again cover a large range of ages with and upper limit governed by their redshift. However, at lower masses galaxies must be increasingly younger to yield a detection. For example at $z\sim5$, if a source is $>100$Myrs old (formation epoch of $z\gtrsim6$) then it must have a mass $>10^{11}$M$_{\odot}$. Therefore, galaxies must either be very young or very massive (or both) to be detected by DES at high redshift.         

\begin{figure*}
\begin{center}
\includegraphics[scale=0.65]{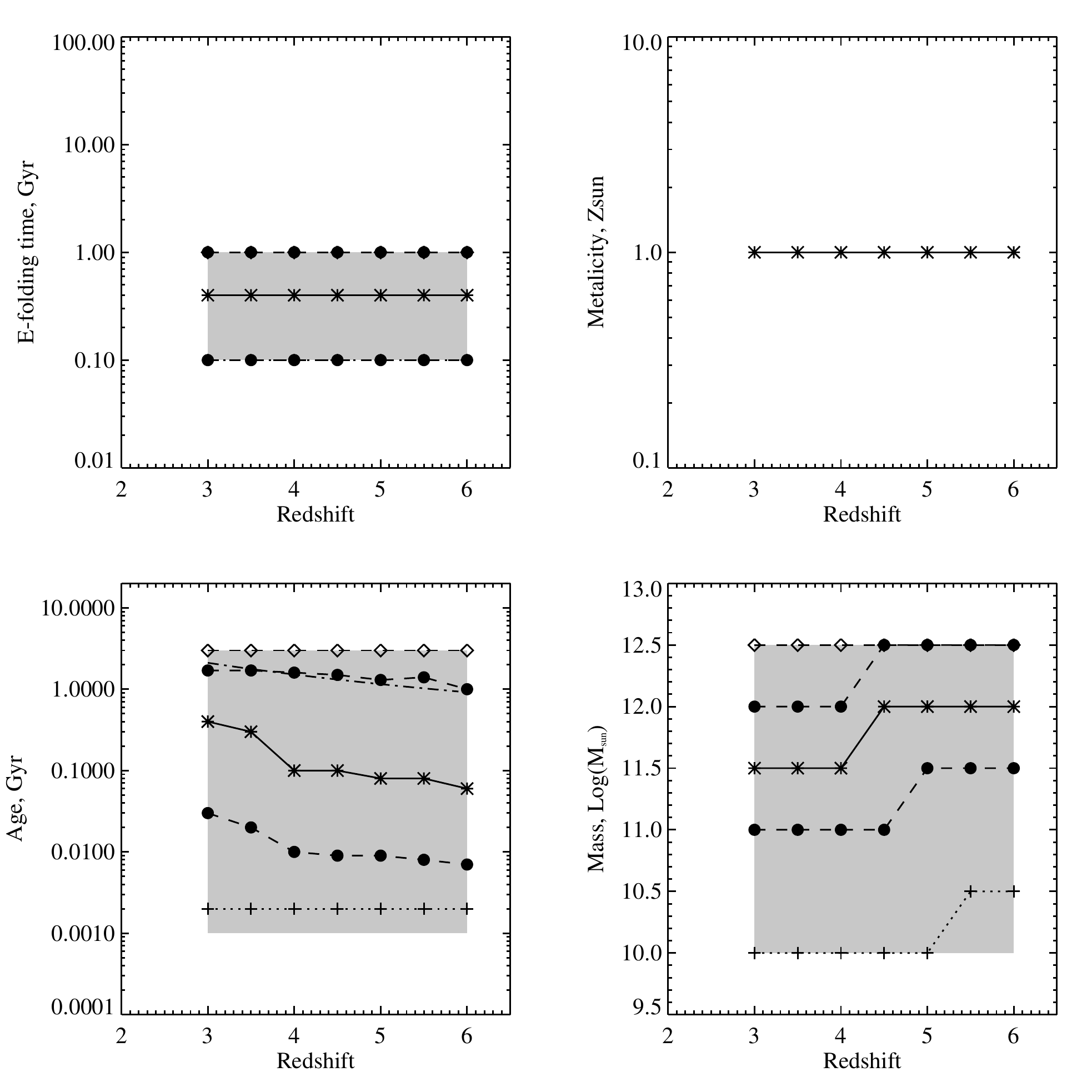}
\caption{Same as Figure \ref{fig:trends} but for the inverted tau models of \citet{Maraston10}. For these models we only consider solar metalicity.}       
\label{fig:trends_invert}
\end{center}
\end{figure*}

Galaxies with exponentially increasing (inverted tau) type star formation histories are also likely to be detected at all redshifts (Figure \ref{fig:trends_invert}), with sources covering a similar range of model parameters.

\begin{figure*}
\begin{center}
\includegraphics[scale=0.8]{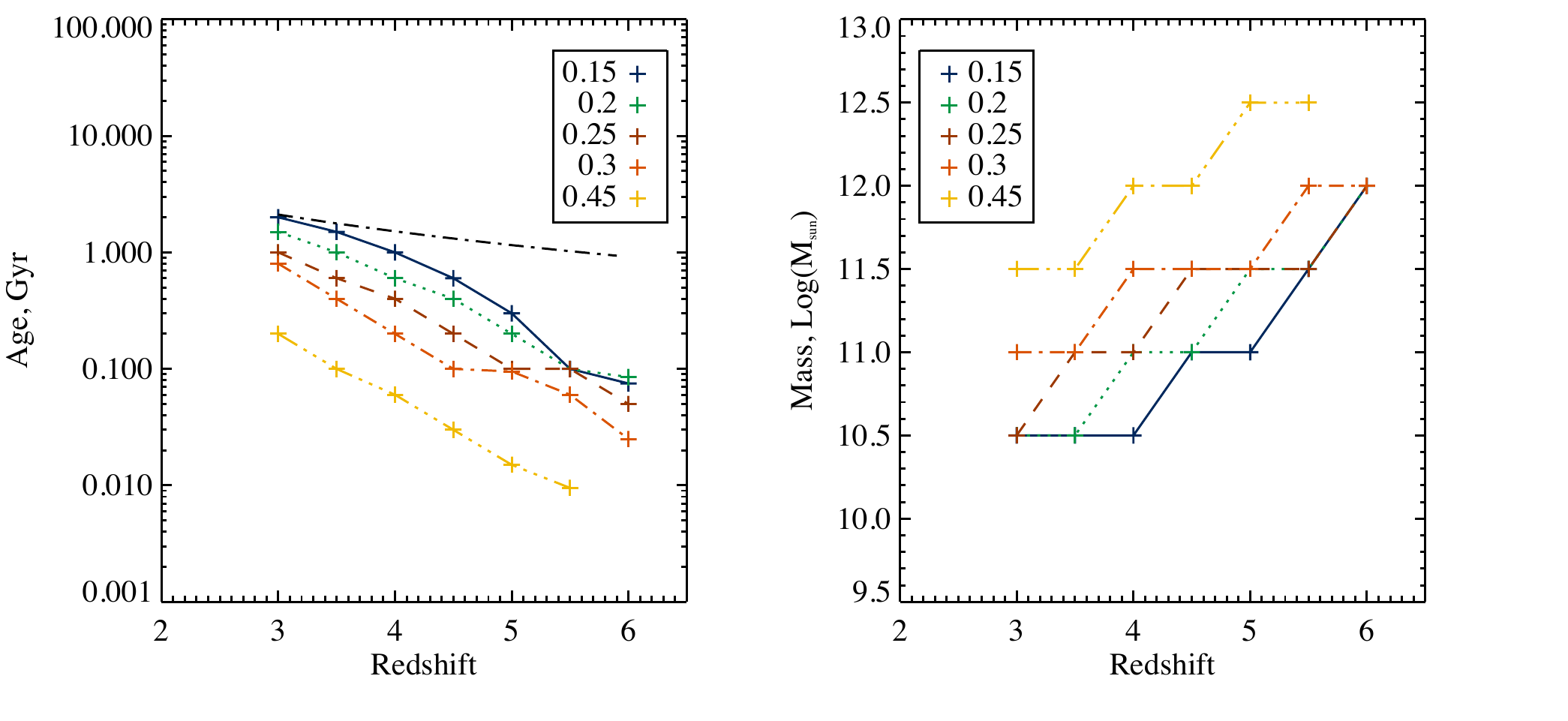}

\caption{The redshift evolution of maximum age and minimum mass of galaxies likely to be detected by DES with a  \citet{Calzetti00} dust extinction model applied for various E(B-V) values. As dust extinctions are increased, sources must be younger and more massive to be detected. Once again the black dot-dashed line represents the age of the Universe at each redshift.}
\label{fig:ex_trends}
\end{center}
\end{figure*}

Thus, modelling potential high redshift galaxies (in a more complex prescription than scaling from lower mass high redshift star-forming galaxies - as discussed in section \ref{sec:nums}) also predicts that DES should detect any massive (log$_{10}$(M$_{*}$/M$_{\odot})\,>\,12.0 $) high redshift systems within its survey volume (for a range of model ages, e-folding times and metallicities). We also find that the majority of detectable objects are very young and therefore their formation epoch will be similar to their detection epoch. It is found that the SSPs models require galaxies with slightly younger ages and lower metallicities than the CSP models in order to be detected by DES, however they generally show similar model properties. For simplicity, in the following discussion we shall only consider the CSP models with exponentially decaying SF.

\subsection{Effects of Dust}
\label{sec:dust}

In order to investigate the effects of dust on the detectability of sources, we apply the \cite{Calzetti00} dust extinction law to all model galaxy spectra and repeat the analysis. In the absence of any information about potential dust extinctions in massive galaxies at high redshift we initially apply the value of E(B-V)=0.15 determined for Lyman break Galaxies (LBG) at $z\sim3$ \citep{Shapley01}.  \cite{Verma07} use SED fitting to a sample of LBGs at $z\sim5$ and find similar dust extinctions (E(B-V)$\sim$0.16), while radio observations of LBGs at $z\sim5$ constrain the dust content of a typical source to be less than one tenth their stellar mass \citep[$\lesssim\,10^8$M$_{\odot}$,][]{Stanway10, Davies12}, suggesting that these low extinction values may also be applicable at $z\sim5$. However, it is interesting to consider the effects of increasing dust extinctions on the properties of detectable galaxies. Extinctions of E(B-V)=0.15-0.45 at 0.05 intervals are applied to the stellar population templates and the process outlined in Section \ref{sec:models} is repeated. We find that for moderate dust extinction, young and massive galaxies are still likely to be detected by DES. Figure \ref{fig:ex_trends} displays the effects of varying dust extinction on the maximum age and minimum mass at which galaxies are detectable. As ought to be expected, the maximum age of galaxies is decreased and the minimum mass is increased as dust extinction is increased. This occurs as brighter more vigorous star-formation is required for a detection. If dust extinctions are increased to E(B-V)\,$>$\,0.6 \cite[similar to those in Sub-mm selected galaxies, $e.g.$][]{Swinbank04} then it is unlikely that any of these galaxies will be detected by DES. However, this result does suggest that older, massive galaxies with moderate dust extinction \citep[such as BzK selected galaxies][]{Daddi05} could potentially be identified in at least one of the DES filters and therefore may be selected with the addition of deep NIR observations such as those from VISTA.

\section{Identifying Massive high-$z$ galaxies in DES}

 \subsection{DES Collaboration Simulated Data}

\label{sec:DES_sim}
  
In preparation for the survey, the DES collaboration has been carrying out detailed catalogue- and image-level low redshift galaxy simulations, as part of annual data challenges that use the simulated data to help develop and test the data management pipelines and science analysis codes \citep{Lin10}. The catalog simulations include dark matter from an N-body simulation box and a galaxy catalogue derived using the ADDGALS method (Wechsler et al., In prep; Busha et al., In prep). ADDGALS is an algorithm for painting galaxies onto dark matter particles in an N-body light-cone simulation by matching galaxy luminosities with local dark matter densities, not dark matter haloes. In addition to the N-body simulation, the algorithm inputs a galaxy luminosity function and a distribution of galaxy colours given luminosity and environment.  Galaxies are assigned to the dark matter on a light-cone, which allows for the direct comparison with galaxy surveys.  The simulation used here extends only to z = 1.3, so it allows us to investigate the contamination from low redshift sources but not directly the high redshift sources of interest.

\begin{figure*}
\begin{center}
\subfigure[$z\sim3$]{\includegraphics[scale=0.4]{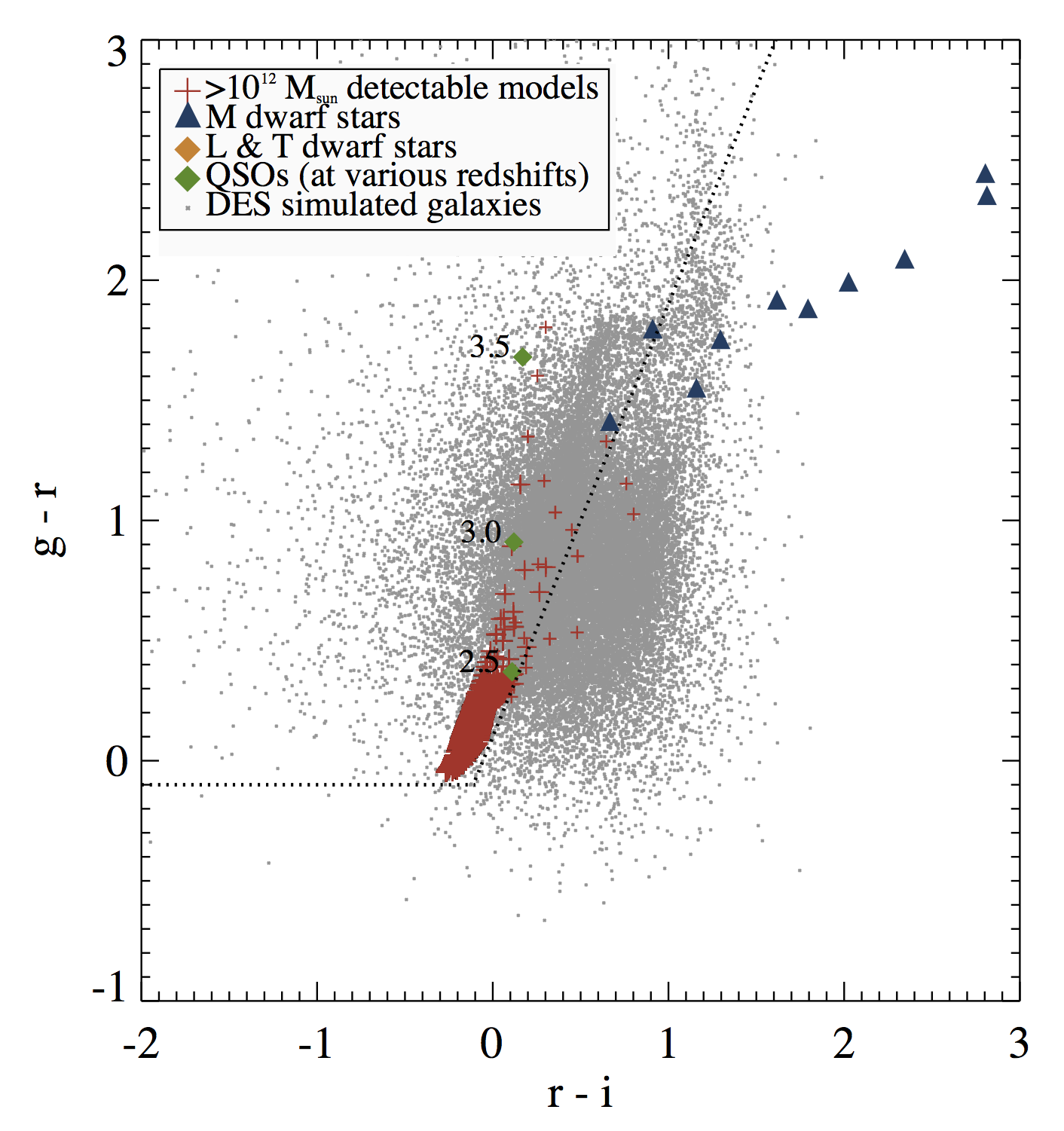}}
\subfigure[$z\sim4$]{\includegraphics[scale=0.4]{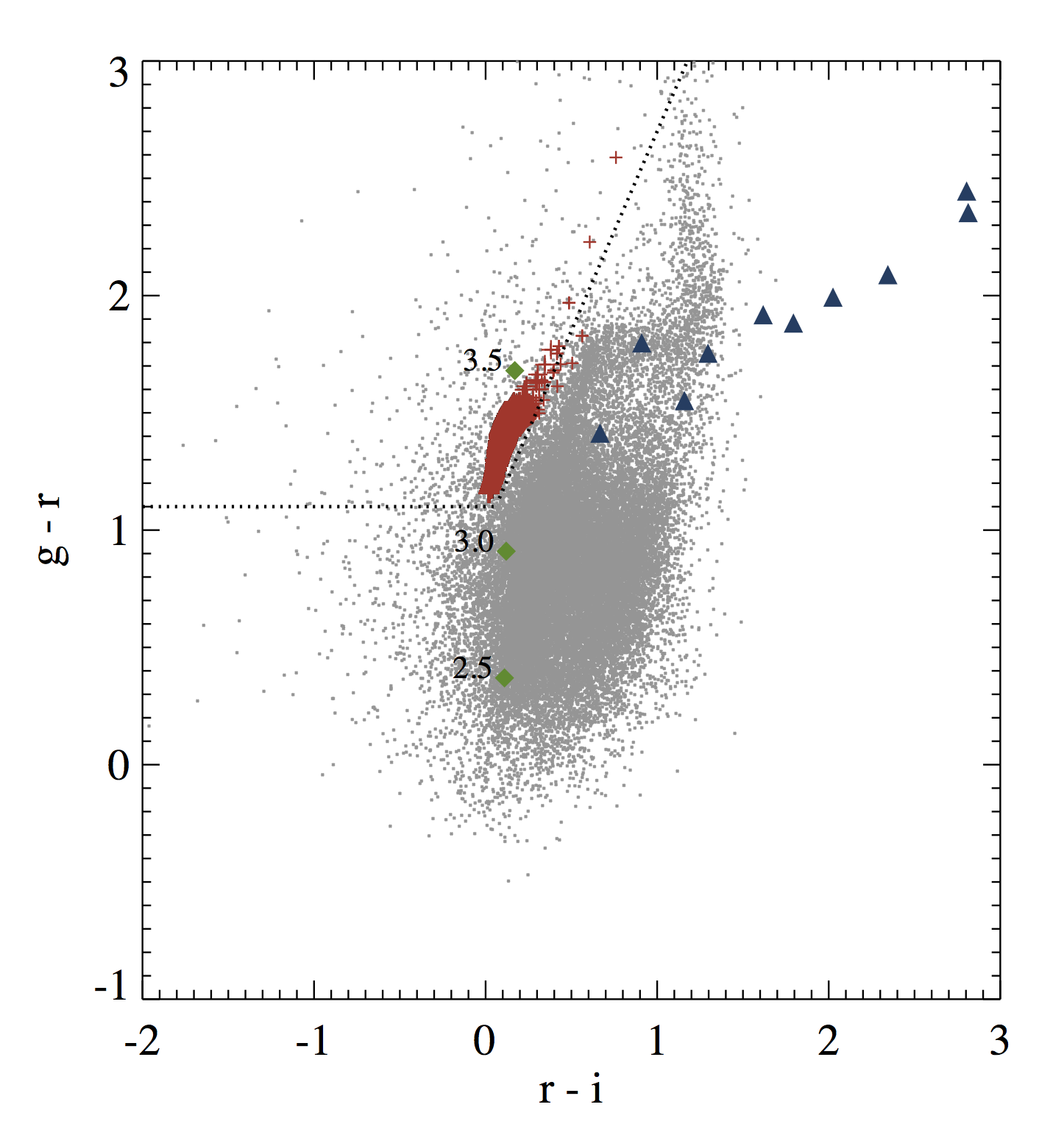}}
\subfigure[$z\sim5$]{\includegraphics[scale=0.4]{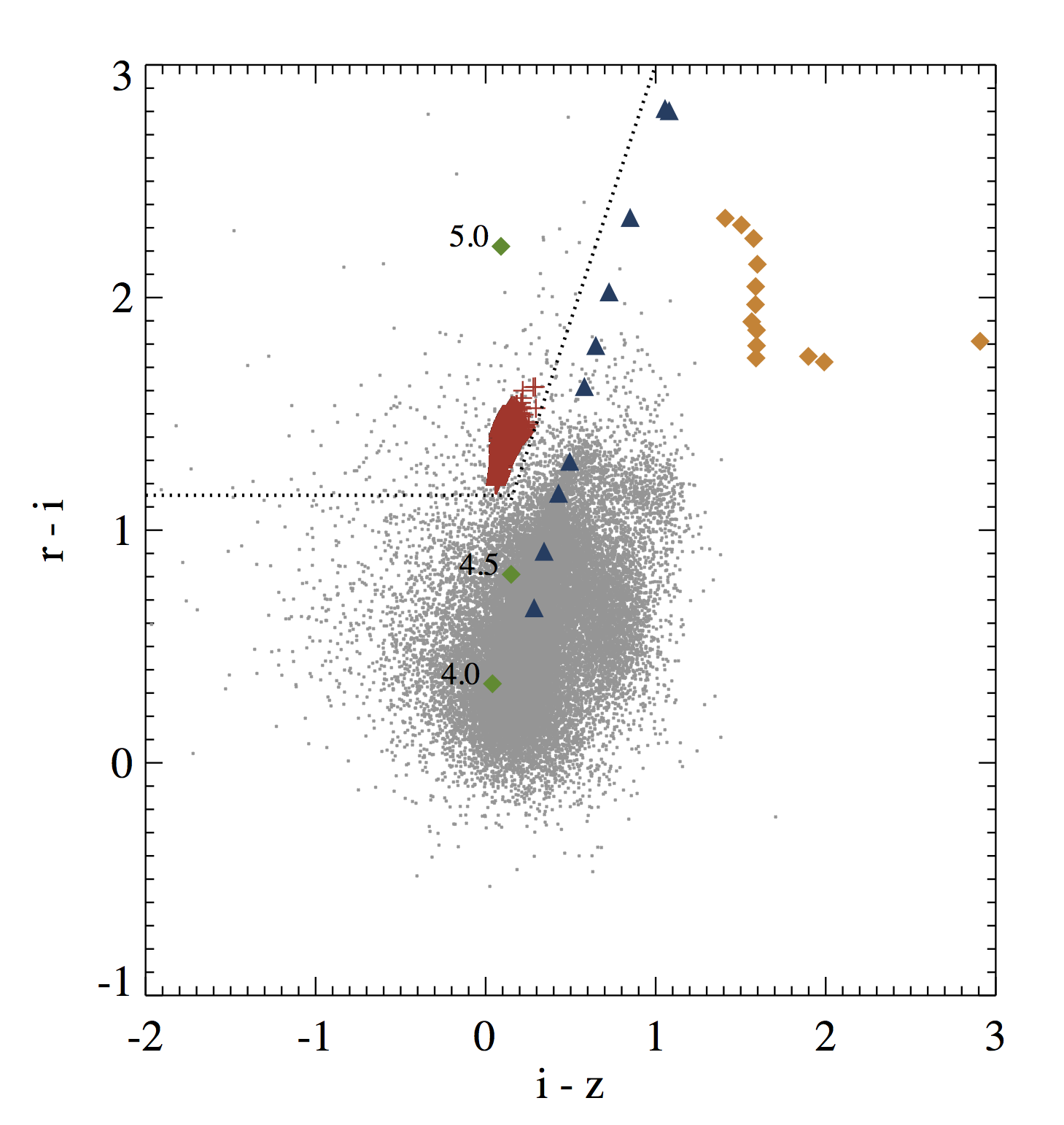}}
\subfigure[$z\sim6$]{\includegraphics[scale=0.4]{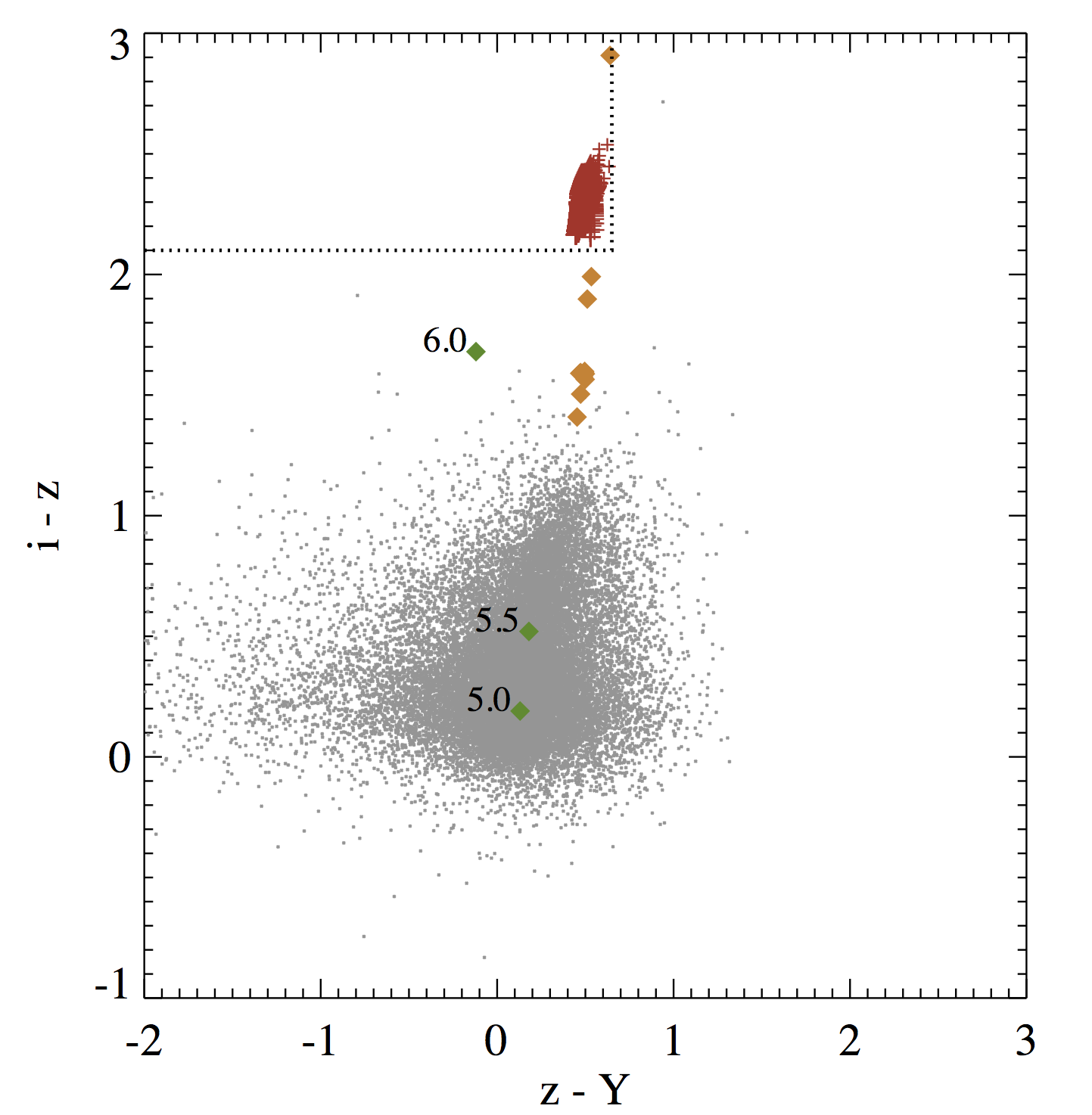}}

\caption{The colour-colour positions of log$_{10}$(M$_{*}$/M$_{\odot}$)$\,>\,12.0$ model galaxies that would be detected in at least one of the DES filters at $z\,=\,3, 4, 5, 6$ (red crosses). A representative random sample of lower redshift galaxies from DES simulated data are displayed as grey points. Blue triangles represent the colours of L and T dwarf stars, while orange diamonds display M dwarf stars. Green diamonds display the colours of QSOs at various redshifts. QSOs display similar colours to possible high-$z$ galaxies and hence can not be differentiated using colour selection criteria. However, if these QSOs are previously unidentified they will also require spectroscopic follow up and hence can be differentiated from high-$z$ galaxies at a later stage. Proposed colour-colour selection regions are bounded by dotted lines. While we do not propose a colour selection for $z\sim3$ galaxies in this paper, in this figure (and subsequent figures) we display a colour selection bounding model sources at $z\sim3$. This highlights how colours selections using the DES filters can not be used to differentiate sources at this epoch from low redshift contaminants.}
\label{fig:col_cols}
\end{center}
\end{figure*}    

 \subsection{Colour-Colour Selection}

Since the mid-90s the most successful method for selecting rest-frame UV-bright galaxies at high redshift has been the Lyman-break technique. This method performs a crude photometric redshift estimation by identifying the large spectral break at 1216\AA\ produced via neutral hydrogen absorption \citep[see][]{Steidel96, Steidel03}. Its successes have proved the practicality of utilising colour selection criteria to identify high-$z$ galaxies \cite[$e.g.$][]{Douglas09,Douglas10, Vanzella09}. 

In order to identify possible massive, rest-frame UV-bright, high redshift galaxies in the DES data set we use a similar colour-colour selection criteria to the Lyman-break technique. The identifiable features in the spectra of target galaxies are shifted to redder bands at higher redshifts, therefore requiring different colour selection at each epoch.

Figure \ref{fig:col_cols} displays colour-colour plots for the detection of possible sources at $z$=3, 4, 5 and 6. Note that the initial colour selections presented in this section are designed to identify relatively `complete' samples of massive high redshift galaxies. Further constraints will be imposed later in this section in order to reduce contamination in our samples and improve the efficiency of any follow up observations - we will not require our samples to be complete only that we can reliably identify a subsample of the most massive sources. The figure shows the colours of mock galaxies from the previous section that are detectable with DES (hence are brighter than the DES limits in at least one of the g, r, i, z or Y bands).  The colours of lower redshift galaxies from the DES simulations (see Section \ref{sec:DES_sim}) are plotted as grey points. We also include the colours of L, T and M dwarf stars and high-$z$ QSOs, which have similar photometric colours to high redshift star-forming galaxies and were not included in the DES simulations.  L and T dwarf stars are plotted as orange diamonds and M dwarfs as blue triangles. Star colours are calculated using template stellar spectra convolved with the DES filter responses - for the L \& T dwarfs \citep{Saumon08,Marley02,Ackerman01} and M dwarfs \citep{Bochanski07}. QSO colours at various redshifts are estimated from the AddQSO algorithms used to predict DES observed colours of high redshift quasars (da Costa et al., In prep)  - plotted as green diamonds. Red crosses represent the $observed$ colours of the model log$_{10}$(M$_{*}$/M$_{\odot}$)$\,>\,12.0$ galaxies which could be detected by DES. Any source which is fainter than the DES limiting magnitudes in either the bluest or reddest band of the three colour selection is set to the DES limiting magnitude. This represent a lower (in the case of the two bluest bands) and upper (in the case of the two reddest bands) limit to the true colour. These source would still be selected in our sample if their true colours were known. For example, a $z\sim5$ source which is undetected in both r-band and z-band may have a lower limit to its r-i colour of 1.0 and upper limit to its i-z colour of 0.0. As our colour selections cover all regions of $r-i>1.0$ and $i-z<0.0$, this source will always match our colour selection irrelevant of its true r-band and z-band magnitudes. Hence, it is included in our selection.          

The simulated massive galaxies of this work can be differentiated from the majority of lower redshift galaxies at all redshifts, except $z\sim3$ where colour selection maybe problematic due to a lack of u-band photometric data. Dwarf stars may provide some contamination sources at $z\sim5$ and 6 \cite[See][for discussion]{Stanway08}. Despite this M dwarf stars in the halo of the Galaxy are typically faint in r, i and z bands. Hence, they are unlikely to be detected at the depths achieved by DES and should not provide a significant level of contamination. QSOs may provide contamination at all redshift. However, if such QSOs are currently undocumented, they will be equally interesting candidates for followup spectroscopic observations. For further discussion of contaminating sources see Section \ref{sec:contam}.

\subsection{Selection Criteria for Relatively Complete Samples}

We propose initial colour selection criteria for the identification of relatively complete samples of massive high redshift galaxies  (dotted lines in Figure \ref{fig:col_cols}) at increasing redshifts (table 1 and figure \ref{fig:col_cols}). The colour selection effectively selects model galaxies and avoids the majority of low redshift objects - except at $z\sim3$ where the high redshift galaxies display similar colours to the DES simulated data and colour selection may be problematic.

\begin{table}
\centering
\large

\caption{The initial colour selection criteria for massive high-$z$ galaxies in DES. For improved colour selection see Table \ref{tab:select2}.}
\begin{tabular}{c c }
\hline
\hline
Redshift & Colour Selection\\
\hline
3& n/a\\
4& $g-r>1.1$ and $g-r>1.7(r-i)+1.0$\\
5& $r-i>1.15$ and $r-i>2.2(i-z)+0.8$\\
6&  $i-z>2.1$ and $z-Y<0.65$\\

\hline
\end{tabular}

\label{tab:select}
\end{table}

\subsubsection{Effects of Redshift}

\begin{figure*}
\begin{center}
\subfigure[$z\sim3$]{\includegraphics[scale=0.4]{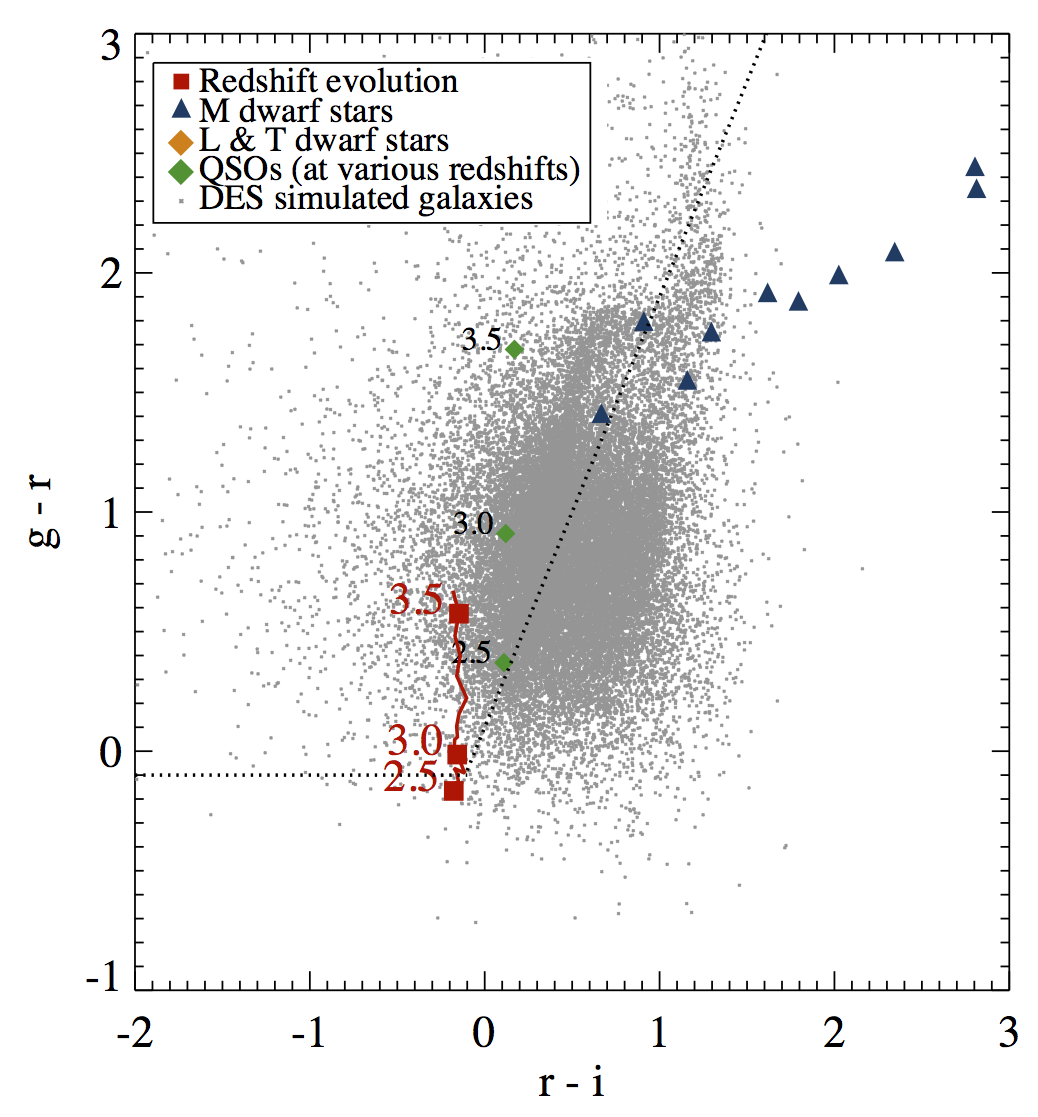}}
\subfigure[$z\sim4$]{\includegraphics[scale=0.4]{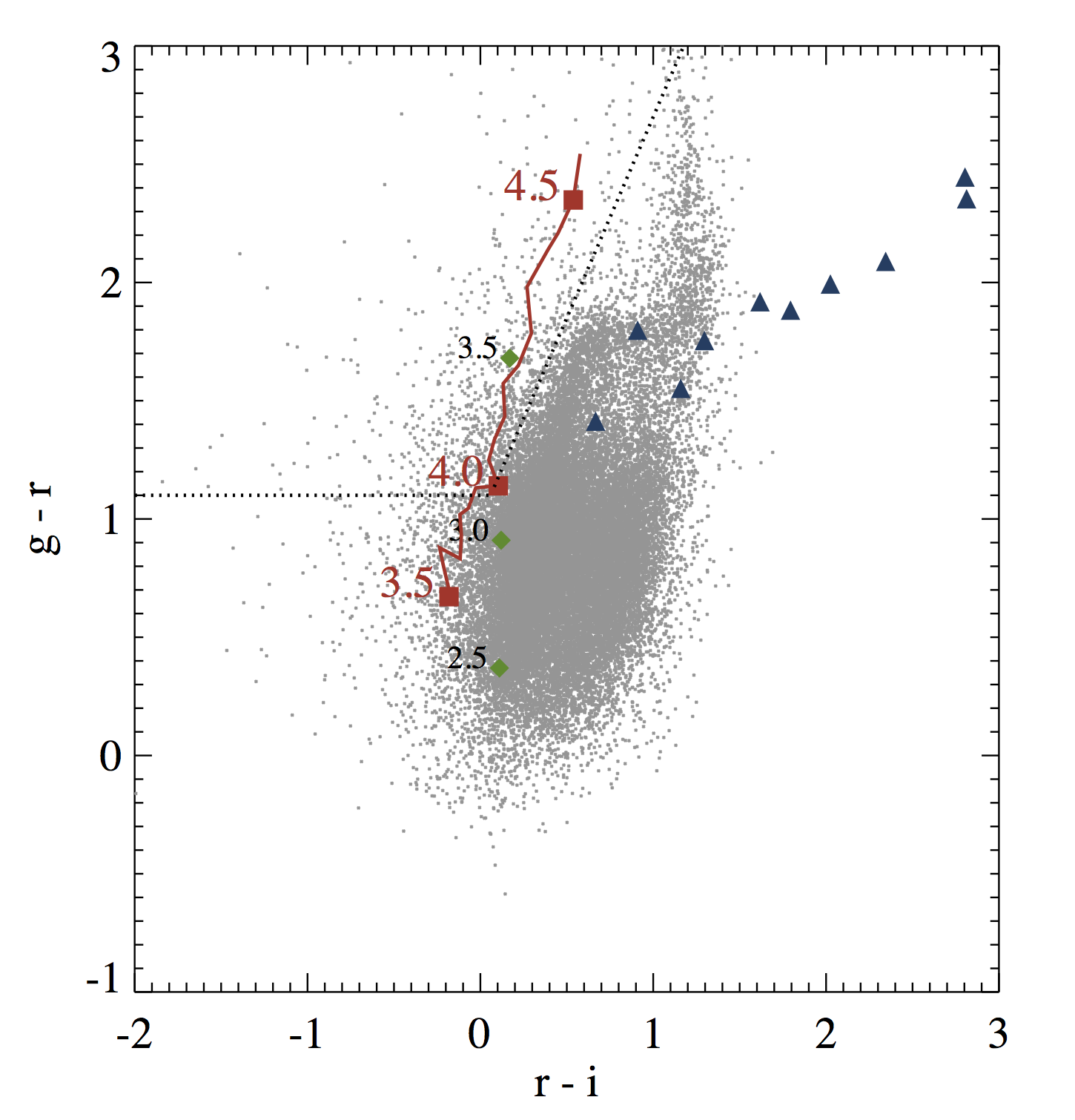}}
\subfigure[$z\sim5$]{\includegraphics[scale=0.4]{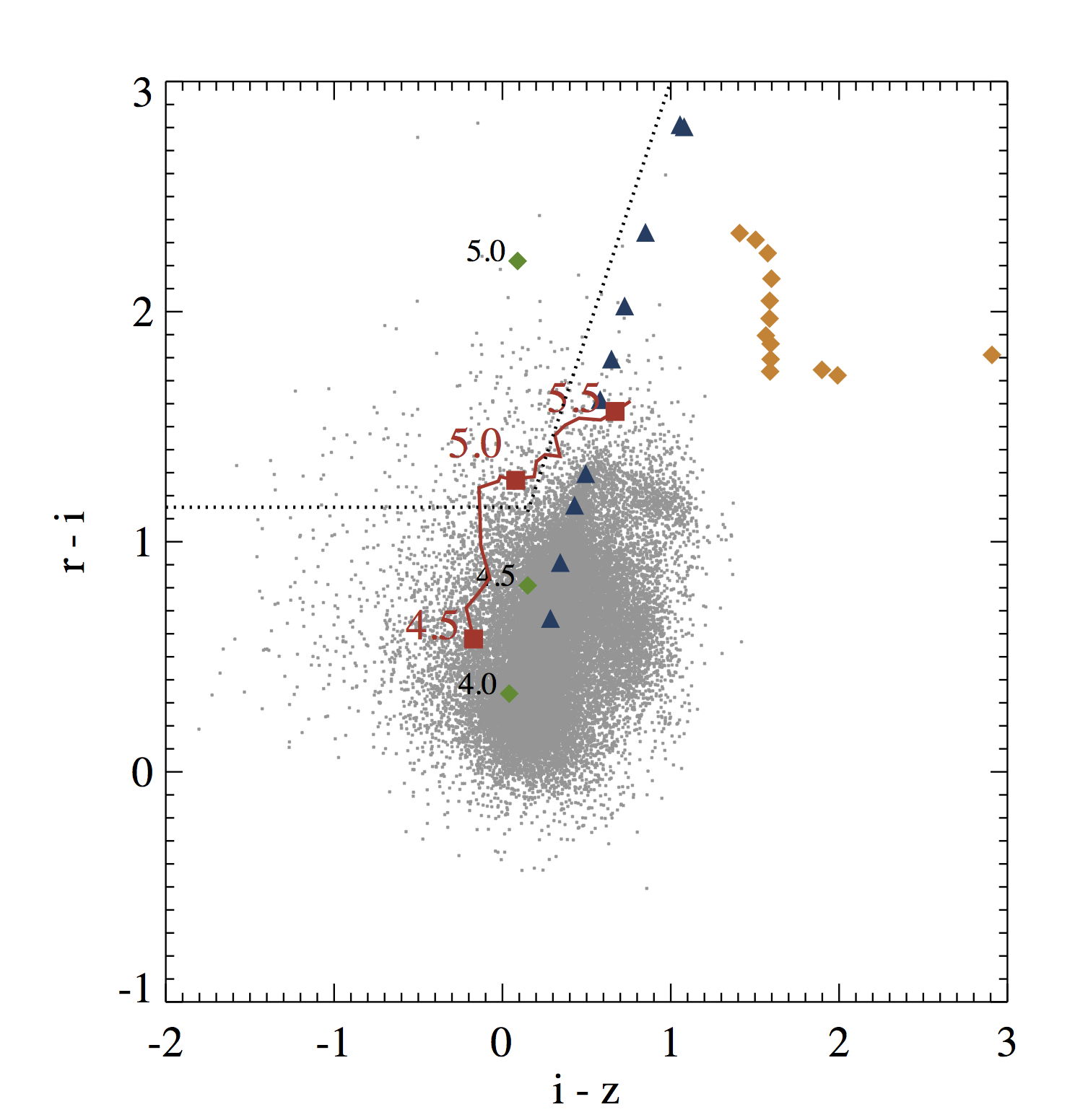}}
\subfigure[$z\sim6$]{\includegraphics[scale=0.4]{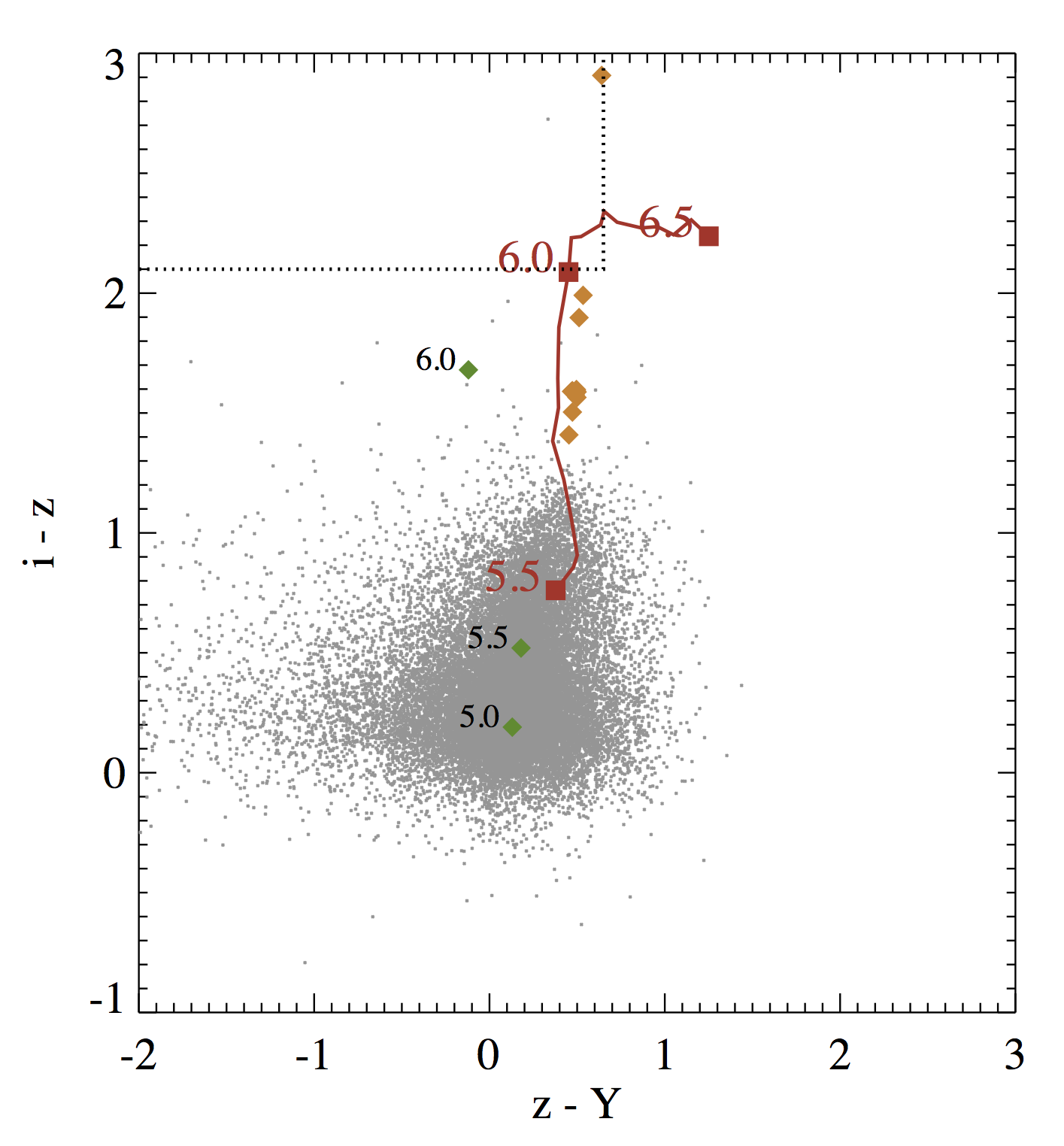}}

\caption{The variation in colour of high redshift galaxies with respect to redshift. Same as Figure \ref{fig:col_cols} but the red line displays the colours of a 10\,Myr old, 0.5 solar metallicity model with an e-folding time scale of 0.5\,Gyr at various redshifts.  The colour selection of massive high redshift galaxies is expanded to include a range of potential redshifts (see table \ref{tab:select2}).  
 }
\label{fig:red_evo}
\end{center}
\end{figure*}

The colour selection criteria outlined in the previous section only identifies galaxies at a particular redshift (those used in the modelling process). However, the colours of high redshift galaxies vary with redshift as the identifiable spectral features move across the filter bands. In Figure \ref{fig:red_evo} we display the change in colours with respect to redshift for a 10\,Myr old, solar metallicity model with an e-folding time scale of 0.5\,Gyr. Clearly, at $z>3$ our initial colour selection (dotted lines in Figure  \ref{fig:red_evo}) will only select objects over a narrow redshift range. Hence, we shall expand our colour selection in order to cover a larger redshift range and thus select a large number of potential high redshift galaxies (see next section and Table \ref{tab:select2}).

\subsubsection{Effects of Dust}
\label{sec:dust2}

\begin{figure*}
\begin{center}
\subfigure[$z\sim3$]{\includegraphics[scale=0.4]{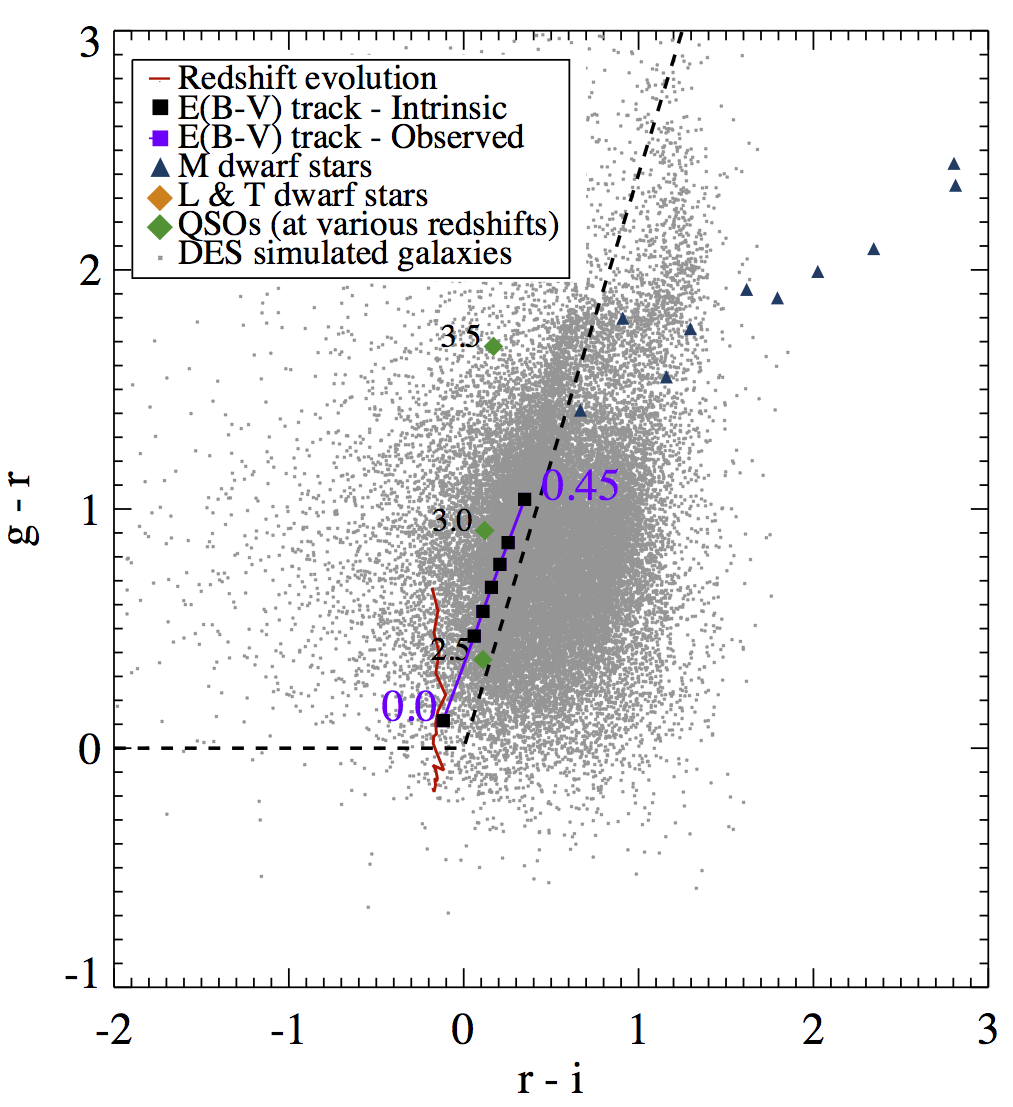}}
\subfigure[$z\sim4$]{\includegraphics[scale=0.4]{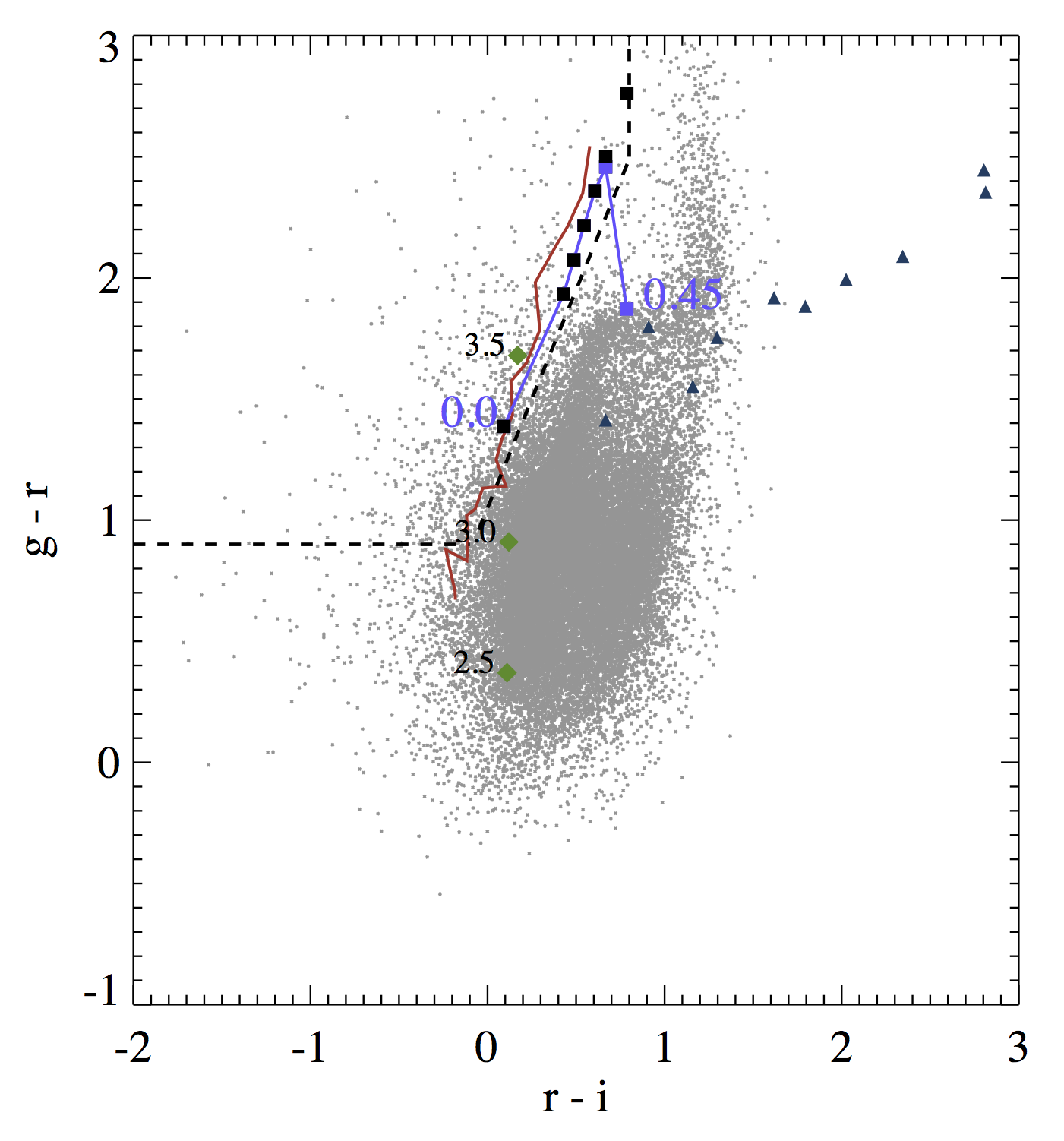}}
\subfigure[$z\sim5$]{\includegraphics[scale=0.4]{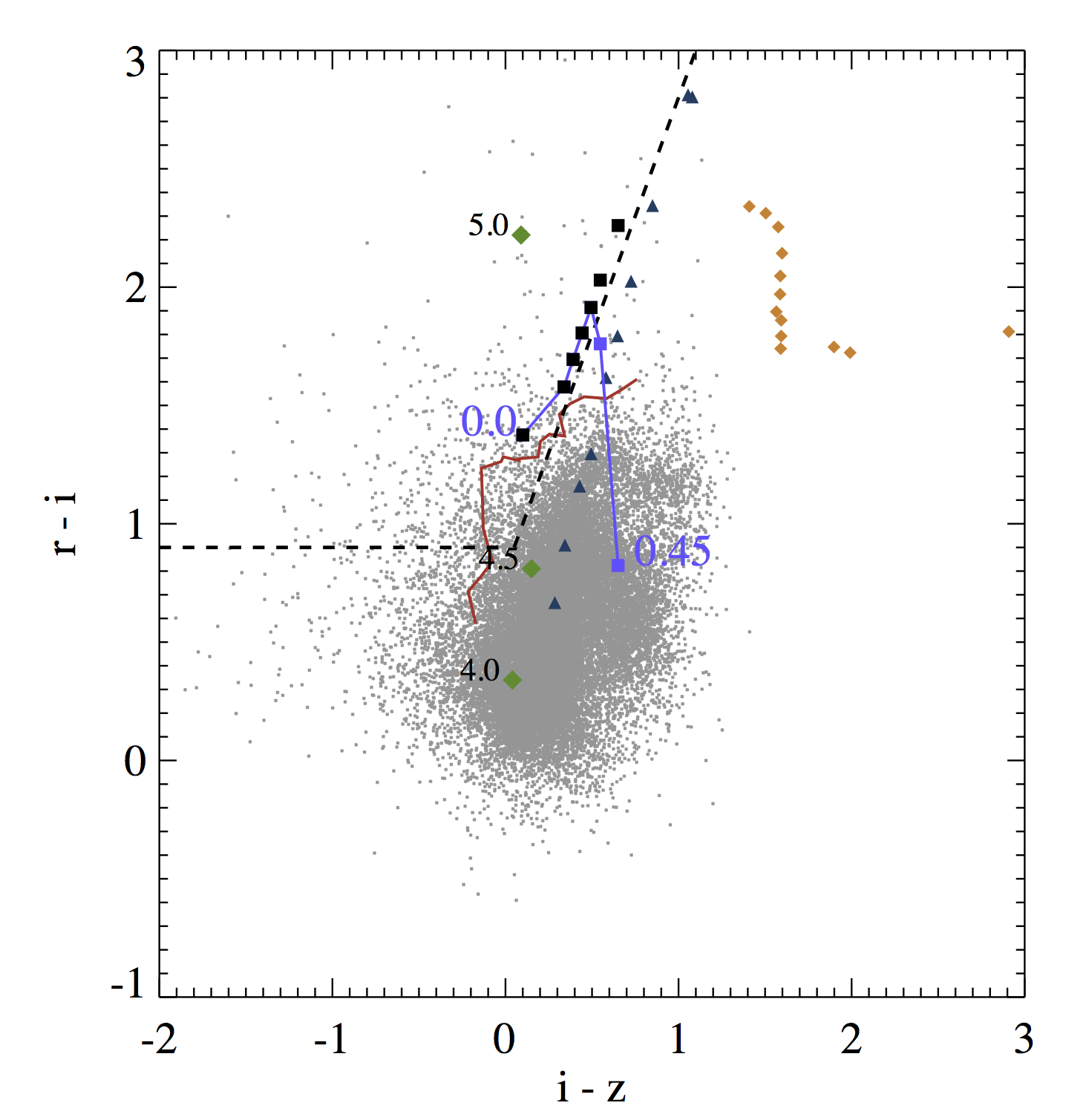}}
\subfigure[$z\sim6$]{\includegraphics[scale=0.4]{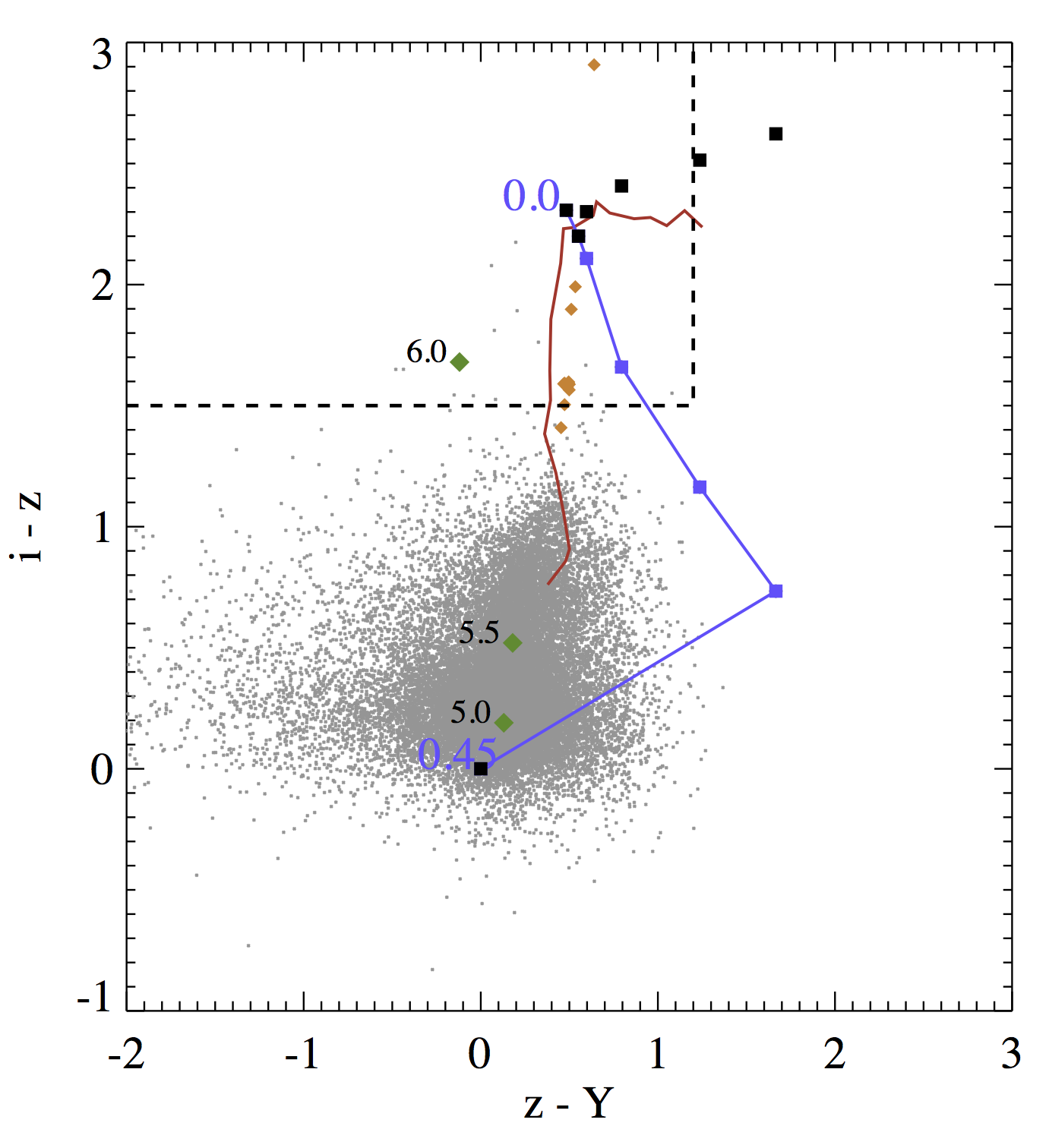}}

\caption{The variation in colour of galaxies likely to be detected by DES as dust extinction is increased. Same as Figure \ref{fig:col_cols} with a \citet{Calzetti00} dust extinction model of varying E(B-V) applied to the model galaxies. The light blue line displays the mean $observed$ (non-detections scaled to DES limiting magnitudes) colour-colour positions of a $10^{12.5}$\,M$_{\odot}$, 10\,Myr old, 0.5 solar metallicity model with an e-folding time scale of 0.5\,Gyr and E(B-V)=0.0, 0.15, 0.2, 0.25, 0.3, 0.35, 0.4, 0.45 (squares). For completeness the true colours of the source (not set to the DES limiting magnitudes) are plotted as black squares. Redshift evolution of non-extincted galaxies from Figure \ref{fig:red_evo} are displayed as the red line. We note that the E(B-V)= 0.45 point at $z\sim6$ represents a non-detection in all bands.}

\label{fig:col_cols_ex_all}
\end{center}
\end{figure*}  

Increasing the dust extinction (as in Section \ref{sec:dust}) will have a significant effect on the colour selection of high-$z$ galaxies. Figure \ref{fig:col_cols_ex_all} displays the effect of applying \cite{Calzetti00} dust extinction of varying E(B-V) on the colours of our model high redshift galaxies. The red line displays a $10^{12.5}$\,M$_{\odot}$, 10\,Myr old, 0.5 solar metallicity model with an e-folding time scale of 0.5\,Gyr with E(B-V)=0.0, 0.15, 0.2, 0.25, 0.3, 0.35, 0.4, 0.45. We once again set the magnitude in any band where the source is fainter than the DES limiting magnitudes to that of the DES limiting magnitude, thus displaying an $observed$ colour - representative of a true DES observation. For completeness we also plot the intrinsic colour of the source (un-corrected to the DES limiting magnitudes) as the black squares. The observed colours at $z=3$ and 4 and initially at $z$=5 and 6 become redder in both colours as dust extinction preferentially attenuates shorter wavelength photons. However, at higher redshifts (and higher extinctions) the flux in the shortest wavelength band (g, r and i at z=4, 5 and 6 respectively) drops below the DES detection threshold. Thus, the \textit{measured} flux in that band (joined by the red solid line) remains roughly constant and the shorter wavelength colour (g-r, r-i and i-z respectively) becomes bluer. This occurs at almost all extinctions at z=6 and is displayed as the turn-over of the red line at $z$=5 and 6.            

At $z\sim3$ the effects of any dust extinction will cause possible high-$z$ galaxies to be indistinguishable from lower redshift contaminants. The identification of galaxies at $z\sim3$ is likely to be problematic without the inclusion of u-band observations as at $z\sim3$ the Lyman spectral break falls below the g-band filter. Therefore, it is unlikely that DES will effectively select galaxies at $z\sim3$. At $z\gtrsim4$ the use of colour selection will identify galaxies of up to moderate extinction (E(B-V)$<0.45$) in the DES sets. However, it must be remembered that these sources have to be increasingly young and/or massive in order to be identified (see Figure \ref{fig:ex_trends}).   

Following this analysis, our colour selection criteria are relaxed to include galaxies of varying dust extinction and redshift (Table \ref{tab:select2} and Figure \ref{fig:col_cols_ex_all}). We note again that these colour selections are designed to be  as complete as possible ($e.g.$ to cover the full range of potential colours of massive high redshift galaxies). As such they will contain a significant number of contaminating sources (see Section \ref{sec:contam}). In practice, the highest priority candidates (in regions with the lowest contamination fractions) will be selected from this colour selection region for initial follow up, with lower priority candidates being targeted if these sources do not yield a detection of a massive high redshift galaxy (see following discussion).

\subsection{Contamination in Complete Colour Selection, Refinements to Reduce Contamination and Follow-up Analysis }
\label{sec:contam}

\begin{figure}
\begin{center}
\includegraphics[scale=0.4]{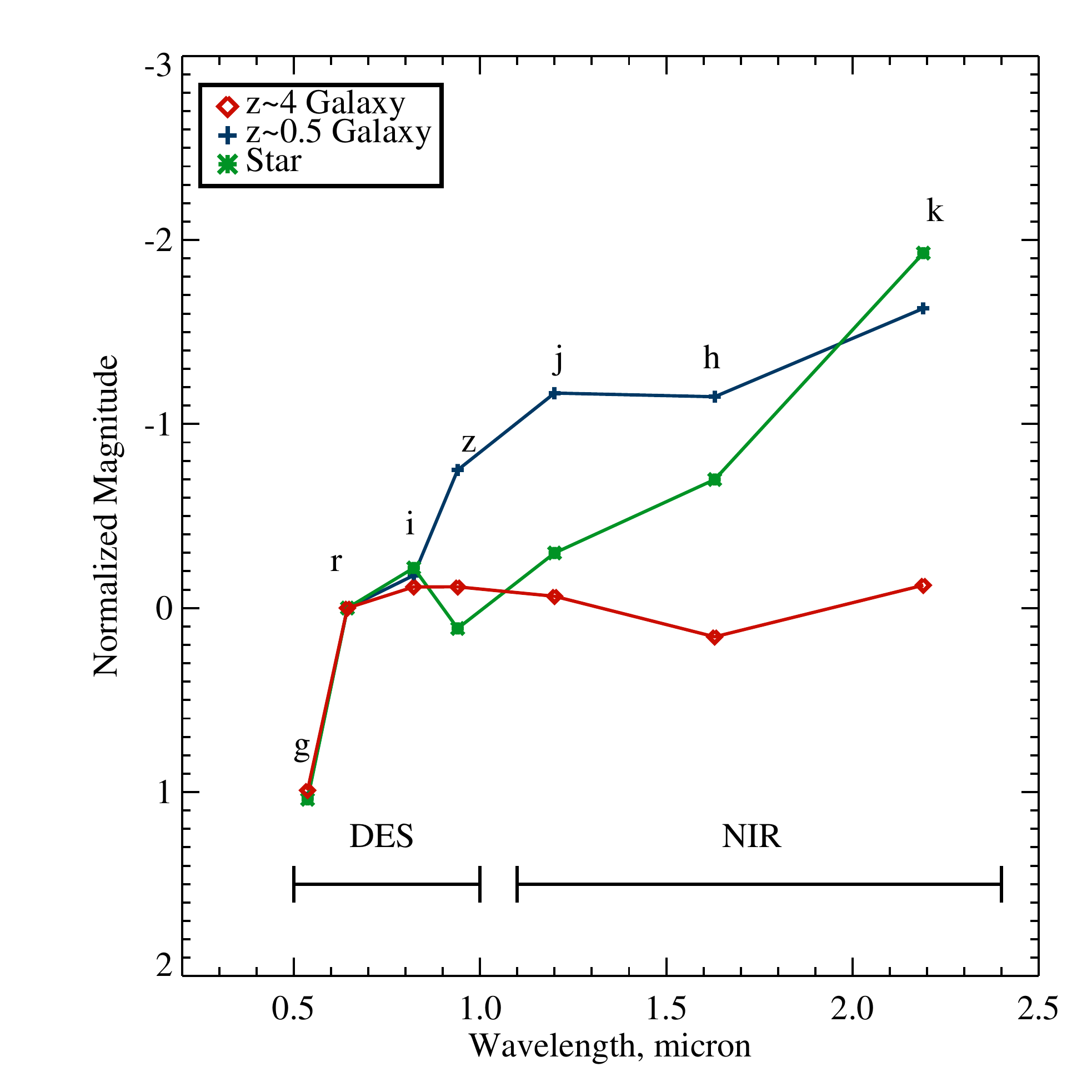}

\caption{Example of a typical optical-IR SED for a $z\sim4$ galaxy (red diamonds), a old, passively evolving $z\sim0.5$ galaxy (blue crosses) and a cool star (green stars) taken from the MUSYC catalogue. Sources are scaled and normalized to the same observed r-band magnitude. Clearly these sources can not be easily differentiated using the DES g, r, i and z colours, and fall in a similar position in colour-colour selection parameter space for $z\sim4$ sources ($i.e.$ the top right panel of Figure \ref{fig:col_cols_ex_all}). However, they can easily be differentiated with the use of NIR observations such as those from VIKING/VHS (For example using i-k colours). We note that while the sources displayed here show different I-z colours, applying an I-z cut to our $z\sim4$ sample does not reduce our potential contamination fraction.  } 
\label{fig:SEDs}
\end{center}
\end{figure}

Using the colour selections outlined above, we will inevitably select a large number of contaminating sources.  Intermediate redshift galaxies (which display similar colours to high redshift galaxies due to their Balmer/4000\AA\ break) and QSOs will be selected in our samples at all redshifts, while L \& T dwarf stars will potentially provide contamination at $z\sim6$ and M dwarf stars at $z\sim5$. Selections including the near-IR data which over-laps with the DES survey area will help remove a significant fraction of the contamination from cool stars and intermediate redshift galaxies \cite[see Figure \ref{fig:SEDs} and ][for discussion of contaminating sources and reducing their effects in colour selected samples at high redshift]{Stanway08}. VHS will cover the majority ($\sim$4500 deg$^{2}$) of the DES survey volume, reaching K-band 3$\sigma$ detection limit of $\sim$20.2, while the deeper VIKING survey's southern field will cover $\sim$500 deg$^{2}$ to a K-band 3$\sigma$ depth of ~21.4. Clearly, these surveys are unlikely to detect our high redshift galaxies (see Section \ref{sec:NIR} and Figure \ref{fig:IR_surveys}), but they will primarily be utilised to rule out any bright contaminating sources with
r-K$\gtrsim$0.5 over almost the entire DES area. Forced photometry on the near infra-red
images using the DES positions will also allow us to push to lower signal-to-noise in 
the near infrared hence potentially reducing contamination and leaving the effective area
scanned for detecting such sources unchanged. If a source which has optical-NIR colours consistent with a high redshift galaxy is serendipitously detected in the NIR observations of VIKING (or even VHS), then it is highly likely to be a very massive source. Therefore, it will be targeted for spectroscopic followup. In addition, priority will be given to potential massive high redshift galaxies which fall in the VIKING region, where contamination fractions will be dramatically lower.

Visual inspection may also be applied to the highest priority targets in order to rule out lower redshift systems.  However, due to the large data sets involved our colour selections for a complete sample are still likely to contain a large number of contaminating sources which can not automatically be ruled out though colour selection or visual inspection. It is difficult to estimate this level of contamination prior to completion of the DES observations. However, we note that the initial DES observations in the VVDS-Deep field (which are still being analysed), contain a similar space density of sources to that predicted from the simulated data discussed in Section \ref{sec:DES_sim}. Comparing to previous studies, typical $z\sim5$ galaxy surveys \citep[$e.g.$][]{Douglas09} obtain $\lesssim40\%$ contamination using similar colour selection to those discussed here. Since studies such as these have deeper detection limits and target fainter high redshift galaxies, they will contain a significantly larger number of both high redshift galaxies and contaminants. Therefore, we may hope to obtain a similar contamination fraction but identify lower number densities of galaxies. While all of the selected sources will not all be the most massive high redshift galaxies, any previously unidentified high redshift sources will also be of interest.

It is possible to investigate the potential level of contamination from photometric scatter into our colour selection regions using the DES simulated data. Applying our colour selections to the simulated data and only selecting sources which are detected in the three bands used for our colour selection at each redshift, excluding the Y-band (as our model massive galaxies are detected in all three bands except Y), we find $\sim$ 265, 220 and 1 simulated sources per deg$^{2}$ in our $z\sim4, 5$ and 6 selections respectively. Clearly this contamination fraction is large and will require further constraints to effectively target massive high redshift systems for follow up studies (see below). However, this contamination does not account for the sources in the DES region which will be ruled out through NIR detections, visual classification and previous studies ($i.e.$ some contaminating sources may have been identified in previous work).   

In addition to the DES simulated data, we may also use a previous obtained deep and large area survey to estimate our potential contamination fraction. Firstly, we consider the Canada-France-Hawaii Telescope Legacy Survey - wide (CFHTLS-W\footnote[1]{Based on observations obtained with MegaPrime/MegaCam, a joint project of CFHT and CEA/DAPNIA, at the Canada-France-Hawaii Telescope (CFHT) which is operated by the National Research Council (NRC) of Canada, the Institut National des Science de l'Univers of the Centre National de la Recherche Scientifique (CNRS) of France, and the University of Hawaii. This work is based in part on data products produced at TERAPIX and the Canadian Astronomy Data Centre as part of the Canada-France-Hawaii Telescope Legacy Survey, a collaborative project of NRC and CNRS.}), which covers an area of 170\,deg$^{2}$ to comparable depths to the predicted DES observations. Hence, this survey will give an indication of the number of sources we may identify in our colour selection region. In order to approximate our colour selection to the CFHTLS-W data, we compare the photometric colours of a flat spectrum source using both the DES and CFHTLS-W broadband filters. Using our $z\sim4$ selection criteria as an example, we find $\sim$400 sources/deg$^{2}$ fall in our selection widow using the CFHTLS-W data. While this level of contamination is high, it tells us little about the fraction of sources in this selection window which are actually high redshift galaxies (which still of great interest and hence, not essential contaminating sources) in comparison to low redshift contaminants.

In order to investigate this further, we utilise the MUltiwavelength Survey by Yale-Chile \citep[MUSYC,][]{Gawiser06}, to estimate the fraction of low-redshift sources which are likely to fall within our colour selections. The MUSYC survey provides deep coverage in 32 (broad and intermediate) bands in the optical and near-IR in the $30^{\prime}\times30^{\prime}$ region of the Extended Chandra Deep Field South (ECDF-S). This detailed photometric coverage allows the determination of accurate photometric redshifts over a large redshift range \cite[see][]{Cardamone10}. Hence, we can select sources from the MUSYC catalogues which are likely to be identified in the DES data, apply similar colour selection and use the accurate photometric redshifts to estimate our contamination fraction. As DES is unlikely to detect $z\sim3$ galaxies and the MUSYC photometric redshift estimates become less accurate/complete at higher redshifts ($z\gtrsim5$), we estimate the contamination fraction at $z\sim4$.    

As with the CFHTLS-W, we compare the photometric colours of a flat spectrum source using both the DES and MUSYC broadband filters. We then scale the MUSYC sample to contain only sources which have a r-band detection at  $<25.0$ (the $5\sigma$ DES limit) and  apply our colour selection criteria. We only include sources with an accurately fit photometric redshift estimate ($<10\%$ error) and compare the fraction of high redshift ($z_{phot}>2.5$) to low redshift ($z_{phot}<2.5$) sources in our colour selection window. Assuming the MUSYC photometric redshifts to be correct we predict a contamination fraction of $\sim38\%$ - similar to that found in spectroscopically confirmed high-redshift LBG searches \citep[$e.g.$][]{Douglas09}.

While the estimated contamination fractions are relatively high (given the exceptionally large DES data sets our colour selections will include a very large number of contaminating sources), it once again does not include any refinements to the candidate selection process involving near-IR data, visual classification, selection of most likely high redshift candidates or SED fitting. Clearly, the colour selections alone will select many candidate high redshift galaxies, but further analysis may be required to target the highest priority candidates for spectroscopic follow up.

\begin{table*}
\centering

\caption{The colour selection criteria for massive high-$z$ galaxies in DES to include objects with dust extinction. We no longer include $z\sim3$ as using the DES data we will not be able to effectively select $z\sim3$ galaxies. We note that these selections are designed to select a relatively $complete$ sample of high redshift sources and are likely to contain a significant contamination fraction. Hence, we will initially target high priority candidates within this region (See Section \ref{sec:contam} for further details).}
\begin{tabular}{c c c c }
\hline
\hline
Redshift & &Colour Selection &\\
\hline
4& $g-r>0.9$ & $g-r>1.8(r-i)+1.05$ & $r-i<0.8$\\
5& $r-i>0.9$  & $r-i>2.0(i-z)+0.9$ & - \\
6&  $i-z>1.5$ & $z-Y<1.2$ & - \\

\hline
\end{tabular}

\label{tab:select2}
\end{table*}

\subsubsection{Refinements to the selection procedure}

\begin{figure}
\begin{center}
\includegraphics[scale=0.4]{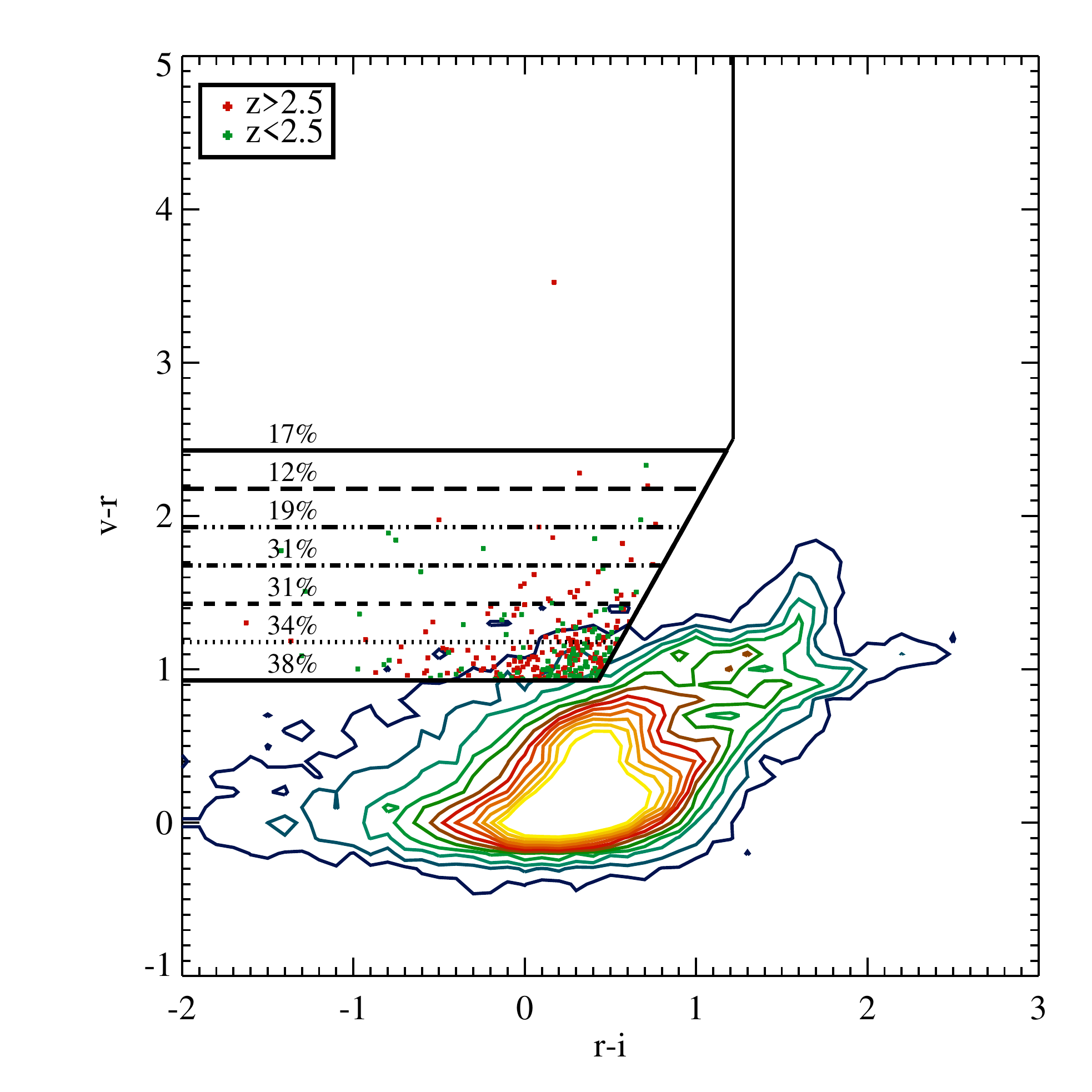}

\caption{Our $z\sim4$ colour selection criteria applied to the MUSYC data (assuming MUSYC v-band magnitudes as a proxy for DES g-band magnitudes and converting our colour selections to the MUSYC filters). Coloured points display sources which would be selected using our colour selection criteria. Red points represent sources with a best fit photometric redshift of $z>2.5$ (high-$z$) and green points a best fit photometric redshift of $z<2.5$ (low-$z$  - contaminants). We vary the v-r cut in the colour selection and estimate the contamination fraction for each cut. v-r cuts are displayed at the horizontal line with the contamination fraction (for sources above that line) displayed on the line. Clearly, at redder v-r colours the contamination fraction is dramatically reduced. Hence, when identifying massive high redshift galaxies in DES we can select the most likely candidates by initially investigating  objects with a very red colours in the two bluest bands in our colour selections (g-r, r-i and i-z at $z\sim4, 5$ and 6 respectively). The bulk of the MUSYC data is displayed as the coloured contours.} 
\label{fig:contam_change}
\end{center}
\end{figure}

In order to highlight how selection of the most likely high redshift candidates can reduce our contamination fraction, we once again use the MUSYC data. Figure \ref{fig:contam_change} displays our $z\sim4$ colour selection criteria applied to the MUSYC data (solid black lines). Here we assume v-band magnitudes as a proxy for DES g-band magnitudes and convert our colour selections to the MUSYC bands. Objects which would be selected as a $z\sim4$ galaxy using our proposed colour selections are displayed as coloured points. Sources with a best fit photometric redshift of $z>2.5$ (high-$z$) are displayed in red and $z<2.5$ (low-$z$ - contaminants) are displayed in green. We vary the v-r cut in our colour selection and calculate the level of contamination for each cut. Horizontal lines display the range of v-r cuts applied in the colour selections. The contamination fraction obtained for the v-r cuts are displayed on each line. Clearly, selecting objects with a very red v-r colour will significantly increase the chances of identifying a true high-redshift source. For example, increasing our g-r colour selection from 0.9 to just 1.4 reduces our $z\sim4$ contamination in the DES simulated data significantly - from $\sim$265 sources per deg$^{2}$ to $\sim$50 sources per deg$^{2}$. Therefore, we refine our colour selections to apply a priority for each source depending on its colour in the two bluest bands of our selection. Table \ref{tab:col_change} displays an example of possible priorities for our $z\sim4$ selection. Clearly, increasing the cut in the g-r colour significantly reduces the contamination fraction and will allow more successful follow up campaigns. We note that the MUSYC contamination fractions are not for the most massive high redshift sources but simply high redshift galaxies. However, by proxy of the DES depths, the majority of high redshift galaxies selected in the DES data will be massive/highly star forming sources. Further priority within these selections will be based on potential K-band detections (although highly unlikely) and, if undetected in the NIR, absolute optical magnitude. Sources with NIR detections and optical-NIR colours consistent with a high redshift galaxy are unlikely to be identified, but if found, are highly likely to be massive sources. In addition, while optical magnitudes may not directly correlate with stellar mass (as instantaneous star-formation dominates rest-frame UV fluxes), on average more massive systems will display larger rest-frame UV fluxes - if the systems is sufficiently young.

\begin{table*}
\centering

\caption{Potential priority assignment within the colour selection window and proposed NIR colour selection in order to reduce contamination fractions. $^{1}$Without NIR colour selection applied. $^{2}$ These numbers contain no other refinements to the selection, such as visual classification or NIR removal of contaminants.}
\begin{tabular}{c c c c c c }
\hline
\hline
Priority & DES selection & DES selection & DES selection &  MUSYC contam$^{1}$ & DES sim contam$^{1,2}$ \\
\hline

1 & $g-r>2.2$ & $g-r>1.7(r-i)+1.0$ & $r-i < 0.8$ & 12$\%$ & $\sim$0.05\,deg$^{-2}$\\
2 & $g-r>1.4$ & $g-r>1.7(r-i)+1.0$ & $r-i < 0.8$ & 31$\%$ & $\sim$50\,deg$^{-2}$\\
3 & $g-r>0.9$ & $g-r>1.7(r-i)+1.0$ & $r-i < 0.8$ & 38$\%$ & $\sim$265\,deg$^{-2}$ \\

\hline
\end{tabular}

\label{tab:col_change}
\end{table*}

\subsection{Summary of source selection}

Therefore in summary, while our initial colour selections are designed to identify the most `complete' sample of high redshift galaxies they will be heavily contaminated (as discussed above). Hence, we will initially follow up the most likely massive high redshift candidates. Our method for targeting potential massive high redshift galaxies for further study will be to initially select sources in the low contamination fraction region of our colour selection window  - those with very large colours in the two bluest bands (g-r, r-i and i-z at $z\sim4, 5$ and 6 respectively). If these high priority sources are not confirmed as massive high redshift galaxies, we will move to colour selection regions with higher contamination fractions. In practice this will limit the number of contaminating sources which are targeted with follow up observations. We shall also prioritise sources with NIR detections in the VHS and VIKING data, and NIR colour consistent with high redshift galaxies (r-K$<$0.5 at $z\sim4$). If non-detected in the NIR data sets, we shall prioritise on absolute optical magnitude. Follow up observations are likely to consist of deep NIR observations to further rule out contaminating sources and constrain galaxy masses, as well as spectroscopic confirmations of redshifts. While spectroscopic campaigns with be relatively time intensive, prioritising candidate selection will allow us to efficiently follow up highly likely high redshift targets. For example, using an 8m class telescope, $5\sigma$ continuum detections of a r=23\,mag galaxy can be obtained in 4h, with likely detections of Lyman-$\alpha$ in just $\sim$1h (assuming typical Lyman-$\alpha$ line strengths from $z\sim4$ sources). Assuming a contamination fraction of $\sim12\%$ for priority 1 targets, we would able to follow up (and confirm) a representative sample of potential massive high redshift galaxies in just tens of hours.               
  
\subsection{Detection in near-IR Surveys}
\label{sec:NIR}

\begin{figure*}
\begin{center}
\includegraphics[scale=0.5]{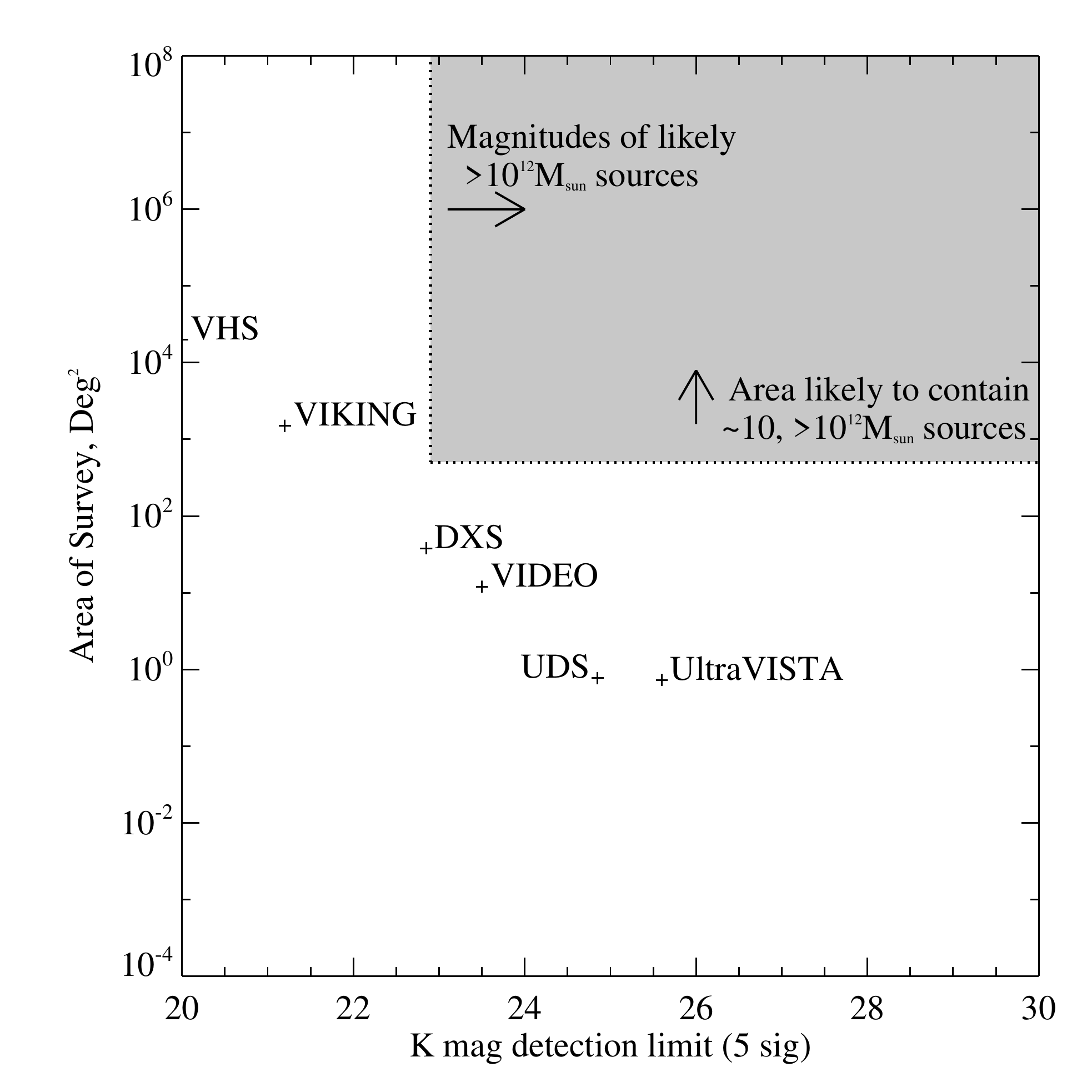}

\caption{A comparison of K-band magnitude limit against area for a sample of surveys. Grey region represent the area of parameter space which is likely to detect M$\,>\,10^{12.0}$M$_{\odot}$ galaxies at $z\,>\,3$. No current or upcoming near-IR survey is likely to detect such sources. Surveys plotted: VISTA Hemisphere Survey (VHS), VISTA Kilo-Degree Infrared Galaxy Survey (VIKING), VISTA Deep Extragalactic Observations Survey (VIDEO), UltraVISTA, UKIDSS Deep Extra Galactic Survey (DXS), UKIDSS Ultra Deep Survey (UDS).} 
\label{fig:IR_surveys}
\end{center}
\end{figure*}

It is also interesting to consider if relatively pristine, massive high-redshift galaxies are likely to be detected in current and upcoming near-IR surveys (independent of a DES detection). Figure \ref{fig:IR_surveys} displays the K-band magnitude limit against coverage for a sample of near-IR surveys in a similar manner to figure \ref{fig:surveys}. Shallow, large area surveys (such as VHS or VIKING) fall at the top left of the plot and deep small field of view surveys (such as VIDEO and UltraVISTA) to the bottom right. The grey shaded region represents the parameter space in which a survey is likely to detect massive high redshift galaxies. The area required to detect massive systems is once again taken from the \cite{Behroozi12} mass function. The magnitude limit is produced by scaling the brightest (K-band) $z\sim5$ star-forming galaxy of \cite{Stark07} to that of a similar 10$^{12.0}$M$_{\odot}$ source (once again assuming luminosity scales directly with mass). This value is representative of a likely K-band magnitude of a 10$^{12.0}$M$_{\odot}$, relatively pristine, star-forming galaxy at high redshift. All surveys fall outside of the parameter space required to detect massive, high redshift systems. Hence, they are unlikely to be detected in current or upcoming near-IR surveys. 

This also poses a potential problem in constraining the mass of our massive high redshift galaxy candidates. The shallow depth of VHS and VIKING surveys suggests that we may not be able to detect even the most massive high redshift galaxies. This is likely to mean that galaxies at $z\sim5$ and 6 will not be detected in the near-IR survey data (as their spectrum remains relatively flat out to $>2400$\AA). As discussed previously, without the addition of near-IR data, determining our candidate sources masses will be problematic. However, at $z\sim4$ this may prove less of a problem as K-band data will observe long ward of the Blamer break and any significant older stellar populations will be easily identifiable. 

Hence, while the deep near-IR surveys will be extremely useful in ruling out low redshift contaminating sources in out sample, further near-IR and spectroscopic follow up may still be required to determine our sources masses and redshifts.

\section{Discussion}

\subsection{Scaled Up Starburst Galaxies}
While many possible progenitors of giant elliptical galaxies have been proposed \cite[See,][for a review]{Renzini06}, each produces difficulties when reconciled with the observable properties of giant elliptical galaxies in high density environments. These systems formed the bulk of their stars in a brief star formation episode at $z\sim5$ \citep{Thomas05, Thomas10} and their high metallicities \citep[\textit{e.g.}][]{Thomas05} suggest they have undergone many successive stages of star formation in a short time \citep{Pozzetti10}. 

A relatively untested method for the production of giant ellipticals is that massive relatively-pristine starburst galaxies formed in the high redshift universe which subsequently passively evolved to the current epoch. These systems may be considered as either a scaled up version of LBGs or sub-mm galaxies without high dust contents (or possibly sub-mm galaxies which have yet to produce their dust). In this model a massive starburst event occurs lasting for a brief period of time until star formation is quenched by some form of feedback ($e.g.$ AGN feedback) or starved by the lack of molecular gas. These galaxies would not require the destruction of dust grains invoked in formation models involving sub-mm galaxies and also do not require the dry mergers of LBG models (see Section \ref{sec:LBGs}). Systems such as this may be produced through amplified star-formation in more normal star-forming galaxies inducing a period of quasi-exponential mass growth \citep{Renzini09}. This star-formation period is then quenched and the system passively evolves.    

The non-detection of these systems to date is unsurprising, given their possible low number densities (if similar to giant ellipticals in the local Universe)  and relatively faint magnitudes. Existing surveys have either been too shallow (\textit{e.g.} SDSS) or have too small a field of view ($e.g.$ SDF) to identify massive, non-dusty, high redshift star forming galaxies. Hence, these possible progenitors of giant elliptical galaxies have been relatively uninvestigated. DES will provide a unique opportunity to identify such systems, should they exist and will provide the first viable candidates for spectroscopic follow up. If massive, non-dusty, high redshift star forming galaxies are identified they will set stringent constraints on galaxy formation models. However, if no convincing candidates are found in the entire DES survey volume then these systems are unlikely to exist.

\subsection{High Redshift Sub-mm Galaxies}

The currently most advocated progenitor of giant elliptical galaxies are high redshift sub-mm selected galaxies \citep[SMGs \textit{e.g}][]{Eales99}. These systems display similar clustering scales to low redshift ellipticals, they are massive \citep[M$_{*}\sim7\times10^{10}$M$_{\odot}$, \textit{e.g.}][]{Hainline10} and have exceptionally large star formation rates \citep[$\gtrsim$\,100\,M$_{\odot}$\,yr$^{-1}$, \textit{e.g.}][]{Tacconi06,Michalowski10} - However, their exact properties are still under debate \citep[$e.g.$][]{Michalowski11}. Therefore, these galaxies do have the potential to form the stellar masses required to produce a massive elliptical galaxy at the current epoch. These intense starbursts may also provide the successive stages of star formation needed to produce large metallicities in just a brief period of time. However, sub-mm galaxies are generally found at lower redshifts ($z\lesssim2.5$) than the formation epoch of giant ellipticals suggested by their stellar histories. While this may be a selection effect \citep[and recently sub-mm sources have been detected at higher redshfits, $e.g.$ ][]{Daddi09, Michalowski10}, it is still difficult to reconcile the observed number densities of higher redshift ($z>2.5$) sub-mm galaxies and low-$z$ giant ellipticals. Could higher redshift precursors to sub-mm galaxies be observed by DES as relatively-pristine, massive, starbursts in the early universe? In addition, sub-mm galaxies (by the nature of their detection) contain large dust contents \citep[M$_{\mathrm{dust}}\sim10^{9}$\,M$_{\odot}$, \textit{e.g.}][]{Michalowski10}, which must be removed or destroyed as they evolve to $z\,=\,0$. The main process for the destruction of dust is thought to be in Supernovae (SNe) driven shocks \citep[\textit{e.g.}][]{Nozawa06}. However, this must not be the case if sub-mm galaxies are the precursor of giant elliptical galaxies, as star formation (and hence SNe driven shocks) must cease after a brief period of time in order to produce the observed stellar histories in low redshift ellipticals. While other methods for the effective destruction of dust grains have been proposed (Clemens et al. 2010), it is still not conclusive as to whether or not sub-mm galaxies can remove their large dust masses as they evolve to $z\,=\,0$. Therefore, while sub-mm galaxies remain likely candidate progenitors to low redshift giant ellipticals it is nonetheless interesting to consider other avenues of massive galaxy formation at high redshift.

\subsection{Small Unobscured Starburst Galaxies}
\label{sec:LBGs}

An alternative scenario is that giant elliptical galaxies are formed from small starburst systems which later merge after their starburst event. High redshift star forming galaxies (such as Lyman Break Galaxies) have been identified in large numbers \citep[$e.g.$][]{Steidel99, Shapley03, Douglas10} and are also undergoing common star burst events. Due to their nature as small independent starbursts, with little molecular gas for subsequent star formation \citep[see ][]{Davies10}, the stars produced in these systems will have a similar, brief, formation epoch. Hence, star formation episodes in LBGs cease without the need to invoke AGN feedback, providing possible explanation of the observed star formation histories in giant ellipticals \citep[$e.g$][]{Gonzalez11}. A caveat to this is the requisite of exclusively $dry$ mergers in forming a giant elliptical galaxy. If the mergers cause subsequent star formation the fossil records of such episodes would be evident in the observed star formation histories of giant ellipticals. In addition, LBGs at $z\sim5$ also display small dust fractions \citep{Stanway10, Davies12} therefore, they do not require dust to be destroyed in their subsequent evolution. While this may make LBGs an attractive candidate for the progenitors of massive galaxies it still does not explain the observed metallicities in giant ellipticals. The formation and dry mergers of small starburst galaxies does not provide the successive star formation episodes required for the production of large metal fractions. In addition, we have not yet found regions which are over-dense in star forming galaxies at high redshift, with which to form a giant elliptical at the current epoch \citep[$e.g.$][]{Douglas10}. However, if DES does not detect the progenitors of massive low-$z$ systems (and sub-mm sources remain undetected at $z\gtrsim5$) then this channel for giant elliptical production gains attraction and helps explain the evolution of the mass-size relation \citep[$e.g.$][]{Oser10}.

\section{Conclusions and future prospects}

 DES will be unprecedented in its ability to provide deep optical coverage over exceptionally large cosmic volumes. While the survey has been designed to constrain dark energy through the study of comparatively low redshift sources ($z\,<\,2$), it will also lead to the identification of rare, massive high redshift galaxies. Such galaxies will have low number densities and will be relatively faint. Hence, they would have evaded current shallow, large area surveys and deep, small area surveys. The unique combination of coverage and depth the DES provides will allow the identification of massive, high redshift star-forming galaxies should they exist, and hence help constrain galaxy formation models. We have modelled possible high redshift galaxies using models galaxies over a broad range of star-formation histories, metallicities, ages and masses, and find that DES is likely to detect massive ($>10^{12.0}$M$_{\odot}$) and relatively pristine (E(B-V)$<$0.45) high redshift galaxies. We have determined colour-colour selection criteria for these galaxies at various redshift intervals which will allow us to differentiate them from lower redshift redshift contaminants. While the existence of massive high redshift, relatively pristine sources will not rule out the currently advocated progenitors of the most low redshift systems ($e.g.$ SMGs or multiple small star-forming galaxies), their identification will help to constrain the earliest stages of massive galaxy formation.      
 
While DES should detect massive high redshift systems (if they are present), it is nonetheless interesting to consider the prospects of future surveys to identify less massive and/or lower number density sources at high redshift. Due to begin in 2020, the Large Synoptic Survey Telescope (LSST) plans to cover 18,000 deg$^{2}$ and reach 27.5\,mag in u, g, r, i, z and Y bands. A survey such as this could potentially observe relatively pristine star-burst galaxies down to stellar masses of $\sim10^{8.5}$M$_{\odot}$ at $z\sim5$ and sources which are three times less numerous than those identifiable by DES. Hence, the LSST will constrain high redshift star-forming galaxies over a wide range of masses and epochs. Looking even further to future projects, ESAs proposed Euclid space telescope \citep{Laureijs11}  will provide a space based extragalactic survey coving 20,000 deg$^{2}$ at NIR wavelengths, allowing the identification of faint, low number density sources at very high redshifts. The potential combination of projects such as DES, and future deep, large area surveys will constrain star-formation over large mass and redshift ranges. This will help develop our ever increasingly detailed picture of galaxy formation and evolution.

\section*{Acknowledgements}

We thank Peter Behroozi for providing us with the data used in Figure 1 and Didier Saumon for the L \& T dwarf star colours provided in Figures 7 - 9. We would also like to thank the anonymous referee for their helpful and insightful comments.

\end{document}